\documentclass[twocolumn,floatfix]{revtex4}%
\usepackage[dvipdfmx]{graphicx}%
\usepackage{amsmath}%
\usepackage{amsfonts}%
\usepackage{amssymb}
\usepackage{bm}
\usepackage[dvipdfmx]{color}
\def\e{{\epsilon}}
\def\k{{ {\bm k} }}

\def\q{{ {\bm q} }}

\def\0{{ {\bm 0} }}

\def\a{{\alpha}}

\def\g{{\gamma}}

\def\r{{ {\bm r} }}

\allowdisplaybreaks[4]

\begin{document}
\title{
Unified mechanism of charge-density-wave and high-$T_{\rm c}$ superconductivity protected from oxygen vacancies in bilayer nickelates
}
\author{
Daisuke Inoue, Youichi Yamakawa, Seiichiro Onari, and Hiroshi Kontani
}
\date{\today }

\begin{abstract}
Unconventional charge and spin density-wave states are commonly observed in bilayer nickelates, drawing considerable attention due to their proximity to high-$T_c$ superconductivity in various phase diagrams.
However, the nature and mechanisms of charge and spin density-waves (DWs) in nickelates remain poorly understood. 
Numerous experiments have reported that the charge-density-wave (CDW) transition temperature $T_{\rm cdw}$ and the spin-density-wave (SDW) transition temperature $T_{\mathrm{sdw}}$ are closely related but distinct.
However, in contrast to these experiments, previous mean-field-type analyses have yielded only a simple SDW phase.
To resolve this key problem, this paper demonstrates that sizable CDW instabilities emerge in proportion to the SDW instability in La$_{3}$Ni$_{2}$O$_{7}$.
This behavior is driven by the paramagnon-interference (PMI) mechanism, which captures important electron correlations beyond mean-field theory.
Therefore, (i) experimental CDW + SDW coexisting state is naturally explained. In addition, (ii) the CDW + SDW fluctuations cooperatively drive high-$T_{\rm c}$ superconductivity. Notably, the predicted $s$-wave SC state is robust against the inner apical O vacancies. Furthermore, (iii) the CDW instability is highly sensitive to the size of the $d_{z^2}$-orbital hole pocket, allowing for the realization of CDW quantum criticality through carrier-doping and pressure application.
We find that the coexistence of charge and spin fluctuations is essential in bilayer nickelates, with both playing a cooperative role in mediating high-$T_{\rm c}$ superconductivity.
\end{abstract}

\address{
Department of Physics, Nagoya University,
Furo-cho, Nagoya 464-8602, Japan. 
}
\sloppy

\maketitle

\section{Introduction}

The recent discovery of high-$T_{\rm c}$ superconductivity in La$_3$Ni$_2$O$_7$ under pressure has generated significant interest. 
The realized superconducting (SC) transition temperature ($T_{\rm c}$) is about 80 K in bulk samples under $P\sim10$ GPa \cite{Sun-Ni-SC, Wang-Ni-SC, Zhang-Ni-SC}, 
while thin-film samples exhibit a $T{\rm c}$ of around 40 K under ambient pressure \cite{Ko-Ni-SC, Zhou-Ni-SC}.
To understand the pairing mechanisms of these high-$T_{\rm c}$ superconductors, which is a central issue of condensed matter physics, the fundamental electron correlations in the normal states have to be clarified.
A key characteristic of these superconductors is the coexistence of charge and spin channel orders in normal states.
In Fe-based superconductors, the nematic and smectic charge-channel orders emerge in many compounds, in addition to the emergence of the spin-density-wave (SDW) states \cite{Kontani-rev2021}.
Based on these observations, both charge-channel and spin-channel fluctuations are naturally expected to mediate the Cooper pairs in various compounds \cite{Prozorov-Fe,Kontani-Fe,Ghigo-SC}.
In cuprate superconductors, short-range or long-range CDW orders are ubiquitously observed near optimally doped compounds, suggesting their significant contribution to the pairing mechanism alongside spin fluctuations \cite{Kontani-rev2021, Ramshaw-SC}.

The phase diagrams of nickelate superconductors La$_{n+1}$Ni$_{n}$O$_{3n+1}$
($n=2$ for double-layer and $n=3$ for tri-layer)
are currently being rapidly elucidated
\cite{Sun-Ni-SC,Wang-Ni-SC,Zhang-Ni-SC,Liu-Ni-opt,Xie-Ni-neutron,Zhang-Ni-phase,Liu-Ni-resistivity,Yang-Ni-ARPES,Cui-Ni-strain,Li-Ni-SC,Wu-Ni-DW,Zhang-triNi-DW,Zhu-triNi-SC,Li-triNi-ARPES,Li-Ni-ARPES,Li-triNi-SC,Yuan-Ni-spin,Zhang-triNi-ARPES,Du-triNi-ARPES,Xu-triNi-DW,Li-triNi-resistivity,Dan-Ni-NMR,Mukuda-Ni-NMR,Chen-Ni-mu+SR,Khasanov-Ni-muSR,Chen-Ni-RIXS,Gupta-Ni-RSXS,Ren-Ni-X-ray,NQR-Okayama}.
For $n=2$ at ambient pressure, double-stripe SDW with ${\bm q}_{\mathrm{s}}\approx(\pi/2,\pm\pi/2)$ appears at $T_{\mathrm{sdw}}\approx150$K, as observed by the $\mu$SR, NMR, and resonant X-ray measurements
\cite{Dan-Ni-NMR,Mukuda-Ni-NMR,Chen-Ni-mu+SR,Khasanov-Ni-muSR,Chen-Ni-RIXS,Gupta-Ni-RSXS,Ren-Ni-X-ray,NQR-Okayama}.
Furthermore, these experiments have observed the emergence of charge density-wave (CDW) orders, with the CDW transition temperature $T_{\rm cdw}$ being comparable to $T_{\rm sdw}$
\cite{Xie-Ni-neutron,Liu-Ni-resistivity,Wu-Ni-DW,Mukuda-Ni-NMR,Chen-Ni-mu+SR,Khasanov-Ni-muSR,Chen-Ni-RIXS,Ren-Ni-X-ray,NQR-Okayama}.
Notably, $T_{\rm cdw}$ seems to strongly depend on the specific samples and their quality.
For instance, recent X-ray measurements report that the double-stripe CDW at ${\bm q}_{\mathrm{c}} \approx (\pi/2, \pm\pi/2)$ emerges at $T_{\mathrm{cdw}} \approx 200$ K \cite{Ren-Ni-X-ray}, which is $\sim50$ K higher than $T_{\mathrm{sdw}}$. 
Also, the relation $T_{\mathrm{cdw}}\approx T_{\mathrm{sdw}}$ is reported in Ref. \cite{NQR-Okayama}.
Notably, the coexistence of CDW and SDW is also a universal feature for $n=3$ \cite{Zhang-triNi-DW}.

In previous theoretical studies,
the importance of the SDW order and fluctuations has been intensively analyzed
\cite{Luo-Ni-SDW,LaBollita-Ni-SDW,Zhang-Ni-SDW,Eremin-Ni-SDW,Gu-Ni-SC,Liu-Ni-SC2,Liu-Ni-SC,Dagotto-Ni-SC,Dagotto-Ni-SC2,Lechermann-Ni-SC}.
Spin-fluctuation-mediated SC states with sign-reversal gap, 
like $d$-wave and $s_{\pm}$-wave states, have been predicted theoretically
\cite{Lechermann-Ni-SC,Liu-Ni-SC2,Heier-Ni-SC,Liu-Ni-SC,Dagotto-Ni-SC,Yang-Ni-SC,Jiang-Ni-SC,Gu-Ni-SC,Dagotto-Ni-SC2,Sakakibara-Ni-SC,Zhan-Ni-SC,Xue-Ni-SC,Tian-Ni-SC,Lu-Ni-SC}.
Also, the normal and SC states have been studied by employing strong coupling approaches such as the dynamical mean-field theory (DMFT)
\cite{Lechermann-Ni-SC,Tian-Ni-SC,Christiansson-Ni-GW_EDMFT,Ouyang-Ni-DMFT,Ryee-Ni-SC}.
However, the origin of the CDW in these nickelates remains unsolved, because strong on-site Coulomb interaction $U$ prevents the CDW order in the scope of mean-field-type analyses.
This issue is crucial for uncovering the pairing mechanism, as the high-$T_{\rm c}$ phase emerges near the CDW quantum-critical-point in the $P$-$T$ phase diagram. 
The key open questions arising from these considerations are:
(i) the origin of the CDW + SDW coexisting states,
(ii) the pairing mechanism driven by CDW + SDW fluctuations, and 
(iii) the key control parameters for density-waves and superconductivity.

To resolve these questions, in this paper,
we demonstrate that both inter- and intra-layer CDW instabilities emerge concurrently with the SDW instability at wavevectors $\bm{q} \approx (\pi/2, \pm\pi/2)$.
This behavior is driven by the paramagnon-interference (PMI) mechanism, which captures important electron correlations beyond mean-field theory.
The PMI mechanism has successfully explained the electric nematic/smectic orders in Fe-based and cuprate superconductors
\cite{Kontani-rev2021}.
We reveal that the vertical bond-order instability in the $d_{z^2}$-orbital develops concurrently with the SDW instability.
Experimental double-stripe CDW/SDW state is naturally explained.
In this mechanism, the CDW instability is highly sensitive to the size of the $d_{z^2}$-orbital hole pocket and is suddenly suppressed by the (electron or hole) carrier-doping.
Importantly, the obtained CDW + SDW fluctuations cooperatively mediate $s$-wave and $d$-wave high-$T_{\rm c}$ superconductivity in nickelates. 
Notably, the predicted $s$-wave SC state is robust against the inner apical O vacancies \cite{vacancy}; see Fig. \ref{fig:fig6} (g).

Theoretically, the bond-order is the modulation of the hopping integrals, originating from the symmetry-breaking in the $\k$-dependent self-energy.
The charge-channel bond-orders have been ubiquitously observed in various strongly correlated metals. 
The PMI mechanism gives natural explanations of various bond-orders observed in Fe-based superconductors \cite{Kontani-rev2021,Onari-SCVC,Onari-Nematic,Yamakawa-Fe}, 
cuprate superconductors \cite{Kontani-rev2021,Tsuchiizu-cuprate,Yamakawa-Fe2}, 
heavy-fermion systems \cite{Tazai-heavy-fermion1,Tazai-heavy-fermion2}, 
twisted bilayer graphene \cite{Onari-TBG}, 
and kagome metal superconductors \cite{Tazai-kagome1,Tazai-kagome2,Tazai-kagome3,Jianxin-kagome1,Jianxin-kagome2}.


\section{Model and formulation}
\begin{figure}[!htb]
\includegraphics[width=.99\linewidth]{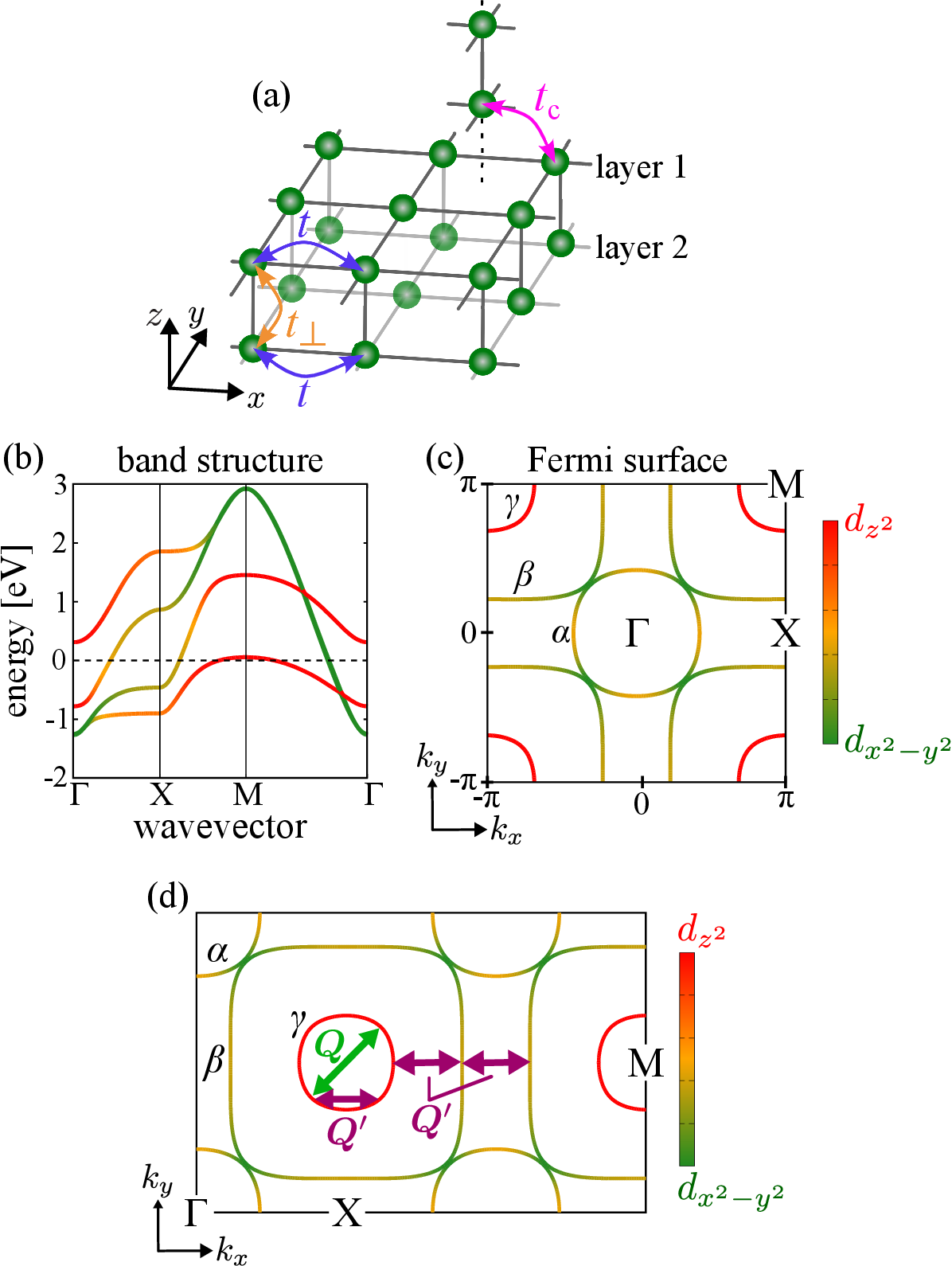}
\caption{
{\bf Tight-binding model for bilayer nickelates}. /
(a) Bilayer lattice structure of $\mathrm{La}_3\mathrm{Ni}_2\mathrm{O}_7$. 
Important inter-$d_{z^2}$-orbital hopping integrals are 
$t_\perp\approx -0.66$, $t\approx -0.11$, and $t_c\approx-0.03$. 
In the present numerical study, we set $t_c=0$. 
(b) Band structure and
(c) FSs for $n=3.0$. The color bar indicates the orbital weight
of $d_{z^2}$- and $d_{x^2-y^2}$-orbitals.
The one-electron pocket is labeled by $\alpha$, 
and two hole pockets are labeled by $\beta$, $\gamma$.
(d) Two main nesting vectors $\bm{Q}$ and $\bm{Q}'$.
$\bm{Q}$ is given by the intra $\gamma$-pocket nesting,
$\bm{Q}'$ is given by the intra $\gamma$-pocket and $\beta$-pocket nestings in addition to the inter ($\gamma$-$\beta$) pocket nesting. 
}
\label{fig:fig1}	
\end{figure}

First, we construct the first-principles bilayer 2-orbital tight-binding model 
of La$_3$Ni$_2$O$_7$ on the square lattice (space group I4/mmm) 
by using WIEN2k and Wannier90 software packages.
The model is based on the experimentally determined lattice parameter 
at $P=29.5$ GPa 
theoretically optimized given in Ref. \cite{Sun-Ni-SC}.
Here, we modified the lattice parameters as $a^*=b^*=(a+b)/2$
for simplicity because the difference between $a$ and $b$ is quite small. 
(From here, we write $a^*$, $b^*$ as $a$, $b$.) 
The model Hamiltonian consists of Ni-ion $d_{z^2}$- and $d_{x^2-y^2}$-orbitals 
for layers 1 and 2 shown in Fig. \ref{fig:fig1} (a).

Figures \ref{fig:fig1} (b) and (c) display 
the band structure and Fermi surface (FS) of the present model for $n=3.0$, 
where the color bar indicates the orbital component. 
In the present numerical study, we set $t_c=0$. 
There are three FSs labeled as $\alpha$, $\beta$, and $\gamma$, 
where $\alpha$-pocket is an electron pocket around $\Gamma$ point, 
and $\beta$, $\gamma$-pockets are hole pockets around M point.
The $\alpha$- and $\beta$-pockets are the mixture of two $d$-orbitals,
while the $\gamma$-pocket is predominantly composed of the $d_{z^2}$-orbital.

In Fig. \ref{fig:fig1} (d),
we explain the main two nesting vectors in the present model.
The nesting ${\bm Q}\approx(\pi/2,\pi/2)$ originates from intra $\gamma$-pocket nesting denoted as the green arrow.
The nesting ${\bm Q}'\approx(\pi/2,0)$ originates from intra $\gamma$-pocket and $\beta$-pocket nestings, in addition to the inter ($\gamma$-$\beta$) pocket nesting. 
In the following numerical study, we use $128\times 128 \ \bm{k}$ meshes and $512-5120$ Matsubara frequencies.
Hereafter, we set $T=5$ meV unless otherwise noted.


In the following, we study the many-body effects due to the $d_{z^2}$-orbital on-site Coulomb interaction (single-orbital $H_U$ model).
This simplification is justified because both the spin susceptibility given by the random-phase-approximation (RPA) and the CDW instability given by the density-wave equation are almost unchanged even if the two-orbital Coulomb interaction model ($d_{x^2-y^2}$- and $d_{z^2}$-orbitals) is analyzed, see the Supplementary Information (SI) A.
The dominance of $d_{z^2}$-orbital has been reported in previous theoretical studies \cite{Luo-Ni-SDW,Eremin-Ni-SDW,Lechermann-Ni-SC,Liu-Ni-SC,Dagotto-Ni-SC,Dagotto-Ni-SC2,Gu-Ni-SC}.
Below, we denote the charge (spin) channel as $b=c \ (s)$.
In the single-orbital $H_U$ model, 
the $b$-channel susceptibility based on the RPA is given as 
$\hat{\chi}^{b}(q) = \hat{\chi}^0 (\hat{1}-\hat{C}^{b}\hat{\chi}^0(q))^{-1}$, 
where $\chi^0_{l_1l_2,l_3l_4}(q)$ is the irreducible susceptibility.
$l_1 - l_4$ ($=1,2$) represent the layer indices, and $q\equiv(\bm{q},i\omega_n)$, 
where $i\omega_n=2n\pi T$. 
Also, $\hat{C}^{b}$ is the bare on-site Coulomb interaction for the $b$-channel:
$C^s_{ll',mm'}=U\delta_{ll'}\delta_{mm'}\delta_{lm}$ and $C^c_{ll',mm'}=-C^s_{ll',mm'}$. 
To express the SDW instability at $\q$, we introduce a factor $\alpha_s(\bm{q})$, 
which is defined as maximum eigenvalue of $\hat{C}^{s}\hat{\chi}^0(\q,0)$. 
Also, the spin Stoner factor is defined as 
$\alpha_{s} \equiv \mathrm{max}_{\bm{q}}\{\alpha_s(\bm{q})\}$.
The magnetic order appears when $\alpha_{s}=1$.

Figure \ref{fig:fig3} (a) shows the obtained intra-layer spin susceptibility $\chi^s_{11,11}(\bm{q},0)$ for $U=1.0$ eV at $T=5$ meV based on the RPA, where the Stoner factor is $\a_s=0.78$. ($\alpha_s=1$ for $U= 1.28$ eV in the RPA.)
The obtained spin susceptibility exhibits quite a broad peak for $|\q|\lesssim\pi/2$, and $\chi^s_{11,11}({\bm Q})$ and $\chi^s_{11,11}({\bm Q}')$ are almost comparable.
Here, the inter-layer susceptibility $\chi^s_{11,22}$ is also enlarged by $U$, becoming comparable to $\chi^{s}_{11,11}$.
Importantly, the relation $\chi^s_{11,11}({\bm Q})\gtrsim\chi^s_{11,11}({\bm Q}')$
is obtained by the RPA based on the ($d_{x^2-y^2}$, $d_{z^2}$)-orbital $H_U$ model, see Fig. \ref{fig:figA1} (b) in the SI A.
Similar results are reported in Refs. \cite{Luo-Ni-SDW,Eremin-Ni-SDW,Lechermann-Ni-SC}.
Figure \ref{fig:fig3} (b) exhibits the $T$-dependences of the spin Stoner factor $\alpha_s$ and $\alpha_s(\bm{Q}')$ for $U=1.05$ eV.
This result means that $\alpha_s(\bm{q})$ is maximum at $\bm{q}=\bm{Q}'$ (${\bm Q}$) at low temperatures ($T\lesssim  10$ meV), while it approaches to $\q\approx{\bm0}$ at higher temperatures.


\section{CDW instability caused by PMI mechanism}

Our main purpose is the theoretical description of the CDW state in La$_3$Ni$_2$O$_7$.
However, the nonmagnetic DW orders cannot be explained within the RPA due to the on-site Coulomb interaction, because $\alpha_s$ is always larger than the charge Stoner factor $\alpha_c$.
In order to explore the CDW order, we have to investigate
beyond-RPA nonlocal electron correlations, described as the vertex corrections (VCs).
The density-wave (DW) equation method enables us to introduce the Aslamazov-Larkin (AL)-type and Maki-Thompson (MT)-type VCs iteratively (see Fig. \ref{fig:fig2} (b) in Method section), which are necessary to preserve the conserving laws.
The derived quantum phase transition from the DW equation automatically minimizes the Luttinger-Ward free energy in Fermi liquids \cite{Tazai-LW}.
This theory has been applied to Fe-based and cuprate superconductors as well as 
kagome metals, and various types of orbital/bond orders in these compounds have been explained satisfactorily \cite{Kontani-rev2021,Onari-TBG,Tazai-kagome1,Tazai-kagome2,Tazai-LW,Yamakawa-BS}.
Here, the orbital/bond orders originate from the AL terms that represent the charge-channel instability caused by the interference between spin-channel fluctuations, called the PMI mechanism
 \cite{Kontani-rev2021,Onari-TBG,Tazai-kagome1,Tazai-kagome2,Tazai-LW,Yamakawa-BS}.
In Method section, we present the explanation for the DW equation.



\begin{figure*}[!htb]
\begin{center}
	\includegraphics[width=.99\linewidth]{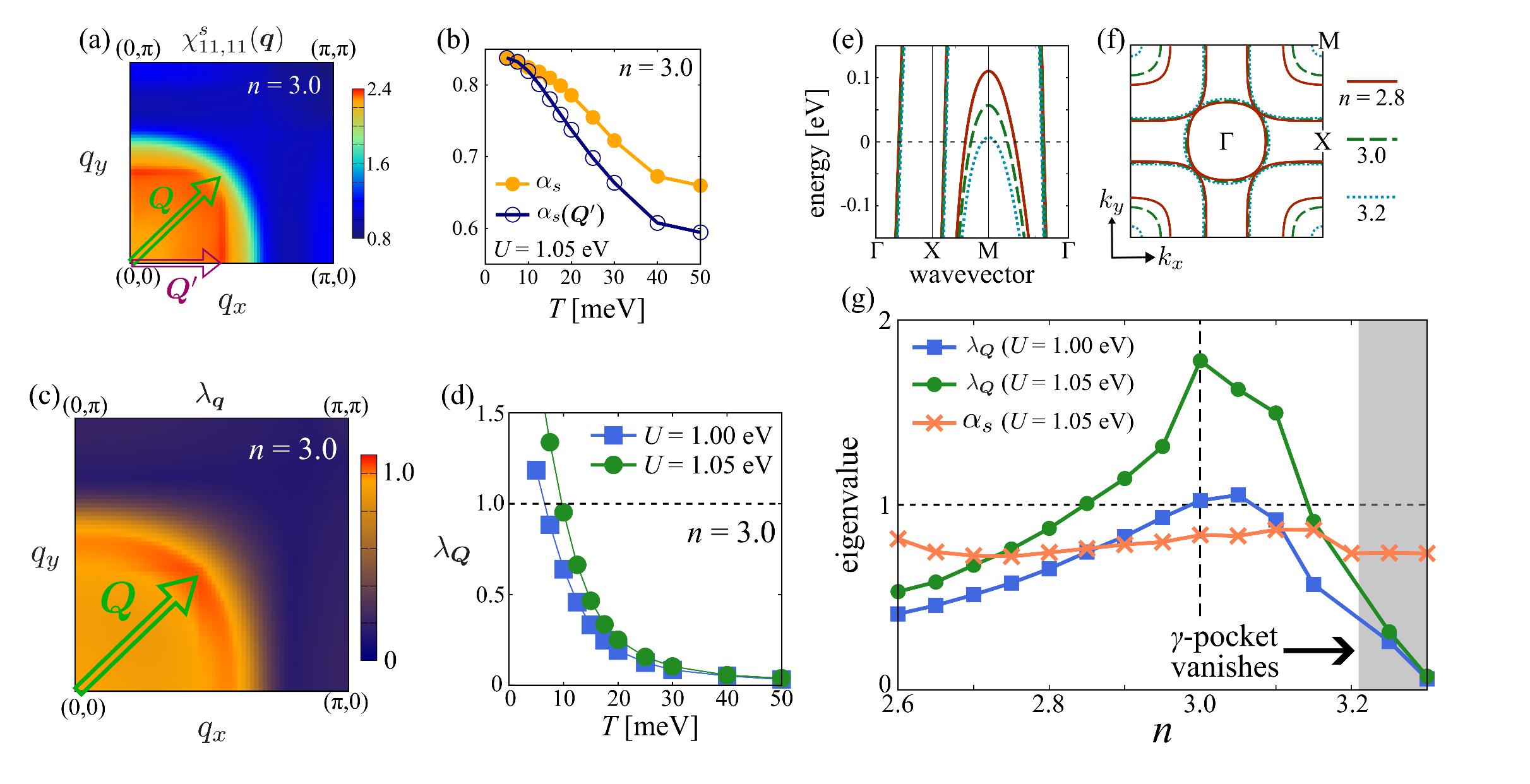}
\end{center}
\caption{
{\bf SDW and CDW instabilities derived from the PMI mechanism}. \
(a) $\bm{q}$-dependences of the intra-layer $\chi^s(\bm{q})$ given by the RPA. 
Maximum peak position of $\chi^{s}(\bm{q})$ is $\bm{q}\approx\bm{Q}'$. 
(b) $T$-dependences of $\alpha_s \equiv \mathrm{max}_{\bm{q}}\{\alpha_s(\bm{q})\}$ and $\alpha_s(\bm{q})$ at $\bm{q}\approx\bm{Q}'$ for $U=1.05$ eV. 
(c) $\bm{q}$-dependences of $\lambda_{\bm{q}}$ for $U=1.0$ eV ($\alpha_s=0.78$) at $T=5$ meV. 
Maximum peak position of $\lambda_{\bm{q}}$ is $\bm{q}\approx\bm{Q}$. 
(d) $T$-dependences of $\lambda_{\bm{q}}$ for $U=$ $1.00$ and $1.05$ eV at $\bm{q}=\bm{Q}$. 
(e) Band structures and (f) FSs for $n=2.6-3.3$. 
$\gamma$-pocket is absent for $n\gtrsim 3.2$. 
(g) Filling-dependences of $\lambda_{\bm{Q}}$ for $U=1.0$, $1.05$ eV, and 
the spin Stoner factor $\alpha_s$ for $U=1.0$ eV. 
Importantly, $n\approx2.6$ is expected in bilayer nickelate thin films \cite{Ni-thin-CDMFT}.
}
\label{fig:fig3}
\end{figure*}

Figure \ref{fig:fig3} (c) shows the obtained eigenvalue of 
the DW equation $\lambda_{\bm{q}}$ for $U=1.0$ eV.
It takes the maximum value at $\bm{q}\approx\bm{Q}$, 
which originates from the incommensurate nesting vector 
in the $\gamma$ pocket (green arrow in Fig. \ref{fig:fig1} (d)).
Here, $\lambda_{\bm{Q}} \approx1.0$ exceeds the SDW instability $\a_s=0.78$.
Importantly, the obtained CDW instability mainly originates from the AL vertex corrections included in the kernel function ${\hat I}$.
In fact, $\lambda_\q\approx0$ in the Hartree-Fock approximation, by dropping all the VCs in ${\hat I}$.
We will discuss the form factor in the next subsection.
Figure \ref{fig:fig3} (d) exhibits the $T$-dependence of $\lambda_{\bm{Q}}$ derived from the DW equation analysis.
For $U=1.05$ eV, $\lambda_{\bm{Q}}$ reaches unity at $T\approx 10$ meV.

Therefore, in the present study, the CDW order without magnetization is realized, and the SDW order will be realized after the CDW transition for larger $U$. 
This result is consistent with the recent X-ray measurement that reports the double-stripe CDW at $\bm{q}_{\mathrm{c}}\approx{\bm Q}$ (at $T_{\mathrm{cdw}}\approx200$ K) in Ref. \cite{Ren-Ni-X-ray}.

Next, we discuss the instabilities of CDW and SDW for both hole-doping and electron-doping.
Changes of the band structures and FSs with carrier-doping are shown in Figs. \ref{fig:fig3} (e) and (f), respectively. 
The $\gamma$-pocket gets smaller by electron-doping and disappears at $n\approx 3.22$ (painted with gray), while $\alpha$ and $\beta$-pockets are almost unchanged.
Figure \ref{fig:fig3} (g) exhibits the $n$-dependence of the CDW eigenvalue $\lambda_{\bm{q}}$ at $\bm{q}=\bm{Q}$ at $T=5$ meV, in the case of $U=1.0$ eV and $U=1.05$ eV.
Also, the orange line denotes $\alpha_s$ for $U=1.05$ eV. 
It is found that $\lambda_{\bm Q}$ is strongly enlarged at $n\approx3.0$, by reflecting the good FS nesting for $\q={\bm Q}$ as shown in Fig. \ref{fig:fig1} (d).
The present result will be useful for understanding the experimental carrier-doping dependence of the CDW phase. 
In the case of hole-doping ($n<3.0$), where $\gamma$-pocket is enlarged, $\lambda_{\bm{Q}}$ moderately decreases.
Thus, the CDW order due to the PMI mechanism is gradually reduced by the hole-doping in La$_3$Ni$_2$O$_7$.
Importantly, $n\approx2.6$ is expected in bilayer nickelate thin films \cite{Ni-thin-CDMFT,Ni-thin-ARPES}.
(The $\q$-dependences of $\chi^s(\q)$ and $\lambda_{\bm{q}}$, and the nesting vectors at $n=2.9$ are shown in Figs. \ref{fig:figA4} (a)-(c) in the SI B.)
In the case of electron-doping ($n> 3.0$), where $\gamma$-pocket shrinks, $\lambda_{\bm{Q}}$ drastically decreases for $n>3.1$.
Because $\a_s$ is less sensitive to carrier-doping,the CDW and SDW instabilities become comparable with carrier (hole or electron) doping.

Experimentally, the SDW order is realized at $T_{\mathrm{sdw}}\approx 150$ K, 
which is common with $\mu$SR, NMR, and X-ray measurements 
\cite{Liu-Ni-resistivity,Dan-Ni-NMR,Mukuda-Ni-NMR,Chen-Ni-mu+SR,Khasanov-Ni-muSR,Chen-Ni-RIXS,Ren-Ni-X-ray,Gupta-Ni-RSXS}. 
In contrast, $T_{\mathrm{cdw}}$ for the charge-channel density-wave appears in a wide temperature range ($T_{\mathrm{cdw}}\approx 110-200$ K) depending on the samples 
\cite{Wang-Ni-SC,Zhan-Ni-SC,Liu-Ni-opt,Liu-Ni-resistivity,Ren-Ni-X-ray}.
One possibility of the reason for this variance of $T_{\mathrm{cdw}}$ 
is self-electron-doping due to the oxygen vacancies. 
In the present numerical study, the maximum peaks of both $\chi^s$ and $\lambda_{\bm{q}}$ approach $\bm{q}=\bm{0}$ due to the reduced $\gamma$-pocket. 
After the $\gamma$-pocket vanishes ($n\approx 3.22$), 
the obtained $\lambda_{\bm{Q}}$ decreases drastically. 

In the SI A, we performed the numerical analysis of the DW equation by considering the two-orbital Coulomb interaction. 
The newly obtained $\lambda_{\bm{q}}$ is very similar to that given by the single $d_{z^2}$-orbital Coulomb interaction model, as we summarized in Figs. \ref{fig:figA1} (b).
This result validates the present analysis based on the single $d_{z^2}$-orbital Coulomb interaction model.

In the present theory, the CDW instability equally develops at $\bm{Q}_1\equiv (1,1)|{\bm Q}|/\sqrt{2}$ and $\bm{Q}_2\equiv(1,-1)|{\bm Q}|/\sqrt{2}$, where ${\bm Q}$ is shown in Fig. \ref{fig:fig1} (d).
Both $\bm{Q}_1$ and $\bm{Q}_2$ are the equivalent nesting vectors in the $\gamma$-pocket. 
Then, the double-$\bm{q}$ CDW ($C_4$ symmetry) or the single-$\bm{q}$ CDW ($C_2$ symmetry) appears at $T=T_{\mathrm{cdw}}$, depending on the fourth-order terms of the Ginzburg-Landau free energy. 
In the SI C, we present a real space picture of the single-$\q$ inter/intra-layer bond-order state, which is suggested in the X-ray experiment \cite{Ren-Ni-X-ray}. 
We also discuss the double-$\bm{q}$ bond-order state in real space in the SI D.





\begin{figure}[!h]
	\includegraphics[width=.99\linewidth]{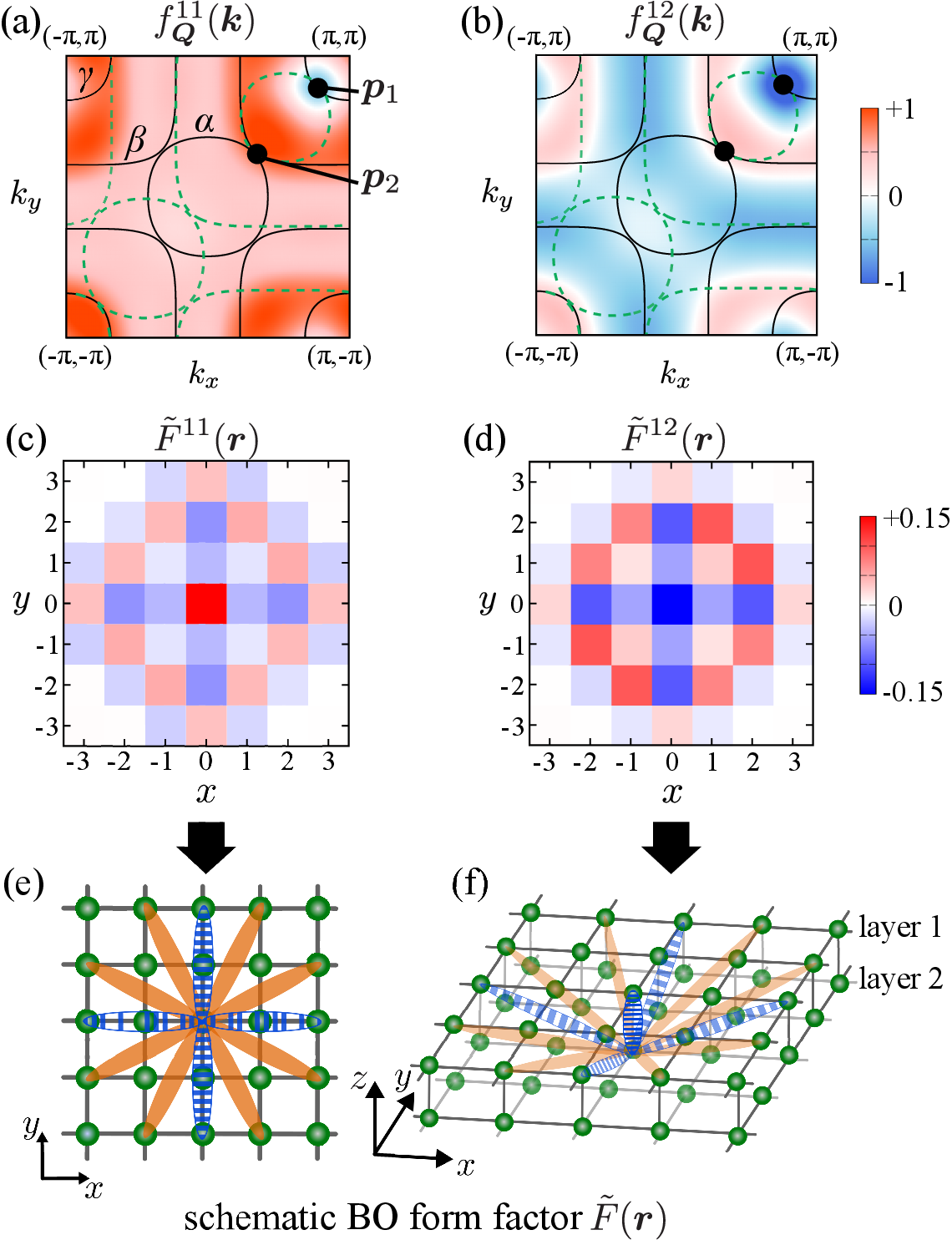}
	\caption{
{\bf Inter-layer and intra-layer form factors for the CDW state}. \
(a,b) Obtained $\k$-space form factors for (a) the intra-layer component ($f_{\bm{q}}^{11}(\bm{k})$) and (b) the inter-layer component ($f_{\bm{q}}^{11}(\bm{k})$).
The black line and green dashed line represent the original FSs and shifted FSs by $\bm{Q}$, respectively.
(c,d) Obtained real-space form factors for (c) the intra-layer component ${\tilde F}^{12}(\bm{r})$ and (d) the inter-layer component ${\tilde F}^{12}(\bm{k})$ at $\bm{q}=\bm{Q}$. 
Here, $\tilde{F}^{11}(\bm{r})=0.54, \ -0.06, \ 0.05$ for ${\bm r}=(0,0)$, $(2,0)$, $(2,1)$, respectively.
Also, $\tilde{F}^{12}(\bm{r})=-0.19, \ -0.10, \ 0.10$ for ${\bm r}=(0,0)$, $(2,0)$, $(2,1)$, respectively.
(e,f) Simplified real-space pictures of the bond-order (BO) form factor: (e) intra-layer form factor and (f) inter-layer form factor. 
Here, orange- and blue-striped bonds denote increased and decreased hopping integrals, respectively.
}
\label{fig:fig4}
\end{figure}

\section{Form factor of inter/intra-layer bond order}

Here, we discuss the obtained CDW order parameter.
Figures \ref{fig:fig4} (a) and (b) show the real part of the obtained intra-layer and inter-layer form factor at $\bm{q}=\bm{Q}$, $f_{\bm{Q}}^{11}(\bm{k})$ and $f_{\bm{Q}}^{12}(\bm{k})$, respectively.
Here, the FSs are denoted by black solid lines, and the shifted FSs by $\bm{Q}$ are drawn by green dashed lines as a reference.
The form factor exhibits large values in magnitudes at $\bm{k}\approx {\bm p}_1$ and ${\bm p}_2$, where the FSs overlap with the shifted FSs.
This is because the particle-hole channel in the DW equation,
given by $\hat{G}(\bm{k})\hat{G}(\bm{k}+\bm{Q})$ 
becomes large for ${\bm k}\approx{\bm p}_m$ ($m=1,2$).	
Thus, the obtained complex ${\bm k}$-dependences of the form factors are not accidental, but a natural consequence of the FS structure of La$_3$Ni$_2$O$_7$.
A more detailed analysis is presented in the SI E. 

To understand the real space DW order with $\bm{q}=\bm{Q}$, 
we perform the Fourier transformation of the form factor: 
\begin{eqnarray}
\tilde{F}^L(\bm{r})&=&\sum_{\bm{k}} F^L_{\bm{q}}(\bm{k}) e^{i\bm{k}\cdot \bm{r}},
\end{eqnarray}
where
$F^L_{\bm{q}}(\bm{k})\equiv f^L_{\bm{q}}(\bm{k}-\bm{\bm{q}/2})$ is the form factor shifted by $\bm{q}/2$, which is convenient for understanding the symmetry of the form factor.
Then, the modulation of the hopping integral $\delta t_{i,j}^L$ is given as 
\begin{equation}
\delta t_{i,j}^L = 2\tilde{F}^L(\bm{r}_i-\bm{r}_j)\mathrm{cos}[\bm{q}/2\cdot (\bm{r}_i+\bm{r}_j)+\psi],	
\label{eqn:hopping_modulation}
\end{equation}
where $\bm{r}_i$ represents the center of a unit cell $i$, and 
$\psi$ is a phase factor. 
Figures \ref{fig:fig4} (c) and (d) show the normalized form factor in real space $\tilde{F}^L(\bm{r})$ for (c) the intra-layer ($L=11$) and (d) the inter-layer ($L=12$) components.
The largest off-site $\tilde{F}^L(\bm{r})$ is the vertical component; $L=12$ and ${\bm r}={\bm 0}$.
In addition, both $|\tilde{F}^{11}(\bm{r})|$ and $|\tilde{F}^{12}(\bm{r})|$ 
take the large value at the third- and fourth-nearest neighbor 
sites, and possess a nearly four-fold symmetry.
Notably, the obtained form factor belongs to $s$-wave symmetry, while it exhibits ``$g$-wave-like'' accidental sign reversals.
Here, the local electron density remains nearly unchanged due to the Hartree term in the kernel function. 
(Local charge modulation induced by the off-site bond-orders would be canceled by the on-site potential $|\tilde{F}^{11}(\bm{r}={\bm0})|$.)
The present intra- and inter-layer bond-order form factors, ${\tilde F}^{ll'}(\r)$, are schematically shown in 
Figs. \ref{fig:fig4} (e) and (f), respectively.
Importantly, the vertical bond-order ($\delta t_{\perp}$) is the dominant order parameter in La$_3$Ni$_2$O$_7$.
We discuss the possible bond-orders in real space derived from the present DW equation analysis in the SI C and SI D.

\begin{figure}[!htb]
	\includegraphics[width=.99\linewidth]{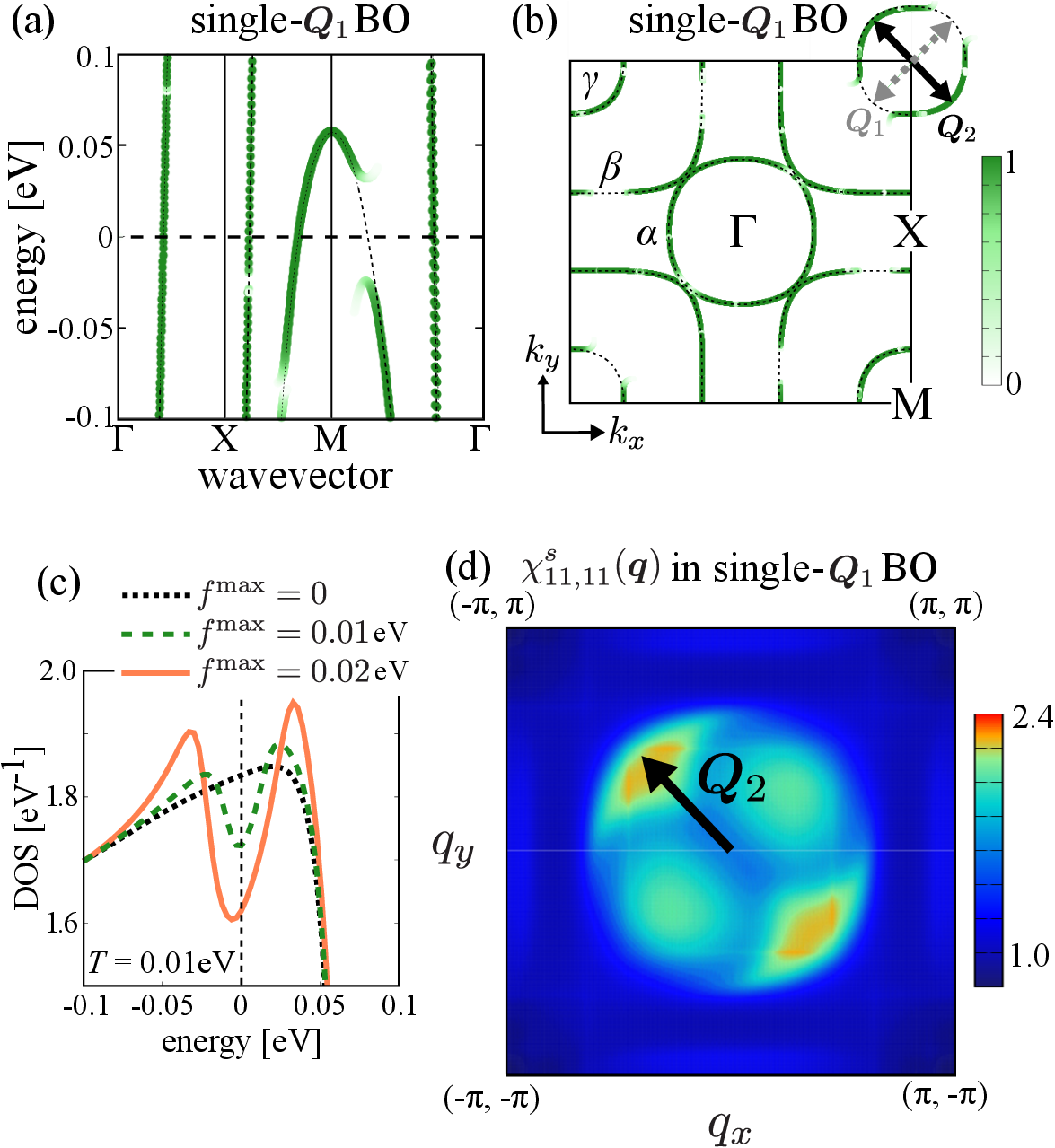}
	\caption{
{\bf Pseudogap formation and double-stripe spin order in the CDW phase}. \
(a) Unfolded band structure and (b) FSs under the single-$\bm{q}$ bond-order 
(green line). 
The black dotted lines denote the band structure and FSs before the CDW order is realized. 
The $\gamma$-pocket almost vanishes. 
(c) DOS under the single-$\bm{q}$ bond-order for $f^{\mathrm{max}}=$0, 0.01, and 0.02 eV. 
(d) Spin susceptibility $\chi^s_{11,11}(\bm{q})$ given by 
the unfolded green function under the single-$\bm{q}$ bond-order with $f^{\rm max}=0.02$ eV. 
	}
\label{fig:fig5}
\end{figure}
Figures \ref{fig:fig5} (a) and (b) show the unfolded band structure and FSs derived from the single-$\bm{q}$ bond order state at wavevector $\bm{q}=\bm{Q}$.
Here, we normalize the form factor $f_{\bm Q}^{ll'}(\k)$ given in Fig. \ref{fig:fig4} (a) and (b) by setting its magnitude as $f^{\mathrm{max}}=0.02$ eV, where $f^{\mathrm{max}}=\mathrm{max}_{\bm{k}}|f_{\bm{Q}}(\bm{k})|$. 
Then, $f^{\mathrm{max}}$ corresponds to the magnitude of the CDW gap $\Delta^{\rm C}$. 
The black dotted lines represent the original band structure and FSs. 
The unfolded band structure exhibits the CDW gap $\Delta^{\mathrm{C}}\sim0.03$ eV 
near the M point. 
Figure \ref{fig:fig5} (c) shows the DOS under the single-$\bm{q}$ 
bond order for $f^{\mathrm{max}}=$ 0, 0.01, and 0.02 eV, respectively. 
The pseudogap opens around 
the Fermi energy $E_{\mathrm{F}}$ for $f^{\mathrm{max}}\neq 0$ eV, 
and the DOS at $E_{\mathrm{F}}$ decreases as $f^{\mathrm{max}}$ increases.

As shown in Fig. \ref{fig:fig5} (b), in the CDW phase, 
nearly half of the $\gamma$-pocket disappears, 
while a wide range of the $\alpha$, $\beta$-pockets remain. 
A large CDW gap opens at Fermi momentum $\bm{k}$, 
where the original FSs and the shifted FSs (by $\bm{Q}_1$) overlap. 
The obtained CDW formation may correspond to the double-stripe CDW with $\q\approx{\bm Q}$ at $T_{\mathrm{cdw}}\approx200$ K reported by recent resonant X-ray measurement \cite{Ren-Ni-X-ray}.
For the double-$\bm{q}$ CDW, the $\gamma$-pocket almost disappears as shown in 
Fig. \ref{fig:figA3} (b) in the SI D. 
Consistently, according to the ARPES study below 160 K \cite{Yang-Ni-ARPES}, 
the $\gamma$-pocket is invisible under the CDW while both $\alpha$- and $\beta$-pockets are clearly observed.

The anomaly of resistivity at $T_{\mathrm{cdw}}$ would be tiny because the $\alpha$- and $\beta$-pockets, which should govern the transport due to their large Fermi velocity, are not gapped by the CDW. 
The present theory may explain the DW-induced disappearance of the $\gamma$-pocket in trilayer nickelate La$_4$Ni$_3$O$_{10}$, whose FSs are very similar to those of Fig. \ref{fig:fig5} (b) \cite{Li-triNi-ARPES}. 

Furthermore, we investigate the SDW order in the single-$\bm{q}$ bond-order. 
In general, the RPA under the CDW state is difficult because the Green function 
$G_{lm}^{\alpha,\beta}$ breaks the translational symmetry, 
where $l$, $m=1-4$ represent the indices of the layers and orbitals, 
and $\alpha$, $\beta$ represents the site in the extended unit cells. 
However, the translational symmetry can be approximately recovered by taking 
the average in the real space called the unfolding 
\cite{Ku-unfold}. 
Now, we perform the RPA calculation under the CDW state using the unfolded Green function $\tilde{G}_{lm}$.
Using this ``unfolded Green function method'', we can efficiently derive the spin susceptibility by taking into account the hybridization gap due to the CDW. 
Figure \ref{fig:fig5} (d) shows the obtained $\chi^s_{11,11}(\bm{q})$ for $U=1.0$ eV under the bond-order with $f^{\rm max}=0.02$ eV.
The maximum peak of $\chi^s$ appears near $\bm{q}\approx\bm{Q}_2$, which is the nesting vector connecting the corner of the $\gamma$-pocket in Fig. \ref{fig:fig5} (b). 
Importantly, the CDW order does not reduce the Stoner factor $\alpha_s$ 
in the case of $\bm{q}_{\mathrm{c}}\neq \bm{q}_{\mathrm{s}}$. 
The predicted cascade transition of the CDW at $\q_{\rm c}\approx \bm{Q}_1$ and the SDW with $\q_{\rm s}\approx \bm{Q}_2$ would be consistent with experiments \cite{Ren-Ni-X-ray,NQR-Okayama}.

In this section, the eigenvalues of the DW equation (in Figs. \ref{fig:fig3} (c) and (d)) are derived from the ``dynamical $f$'' analysis, in which the $\epsilon_n$-dependent form factor $f_{\bm{q}}^L(\bm{k},\epsilon_n)$ is solved accurately.
To avoid the difficulties of numerical analytic continuation ($i\e_n\rightarrow \e+i\delta$), the form factors (in Figs. \ref{fig:fig4} (a), (b)) are derived from the ``static $f$'' analysis.
The eigenvalue given by the static $f$ analysis is about 30\% larger than that by the dynamical $f$ analysis.

\section{Superconductivity by CDW + SDW fluctuations}

\begin{figure*}[!htb]
	\includegraphics[width=.99\linewidth]{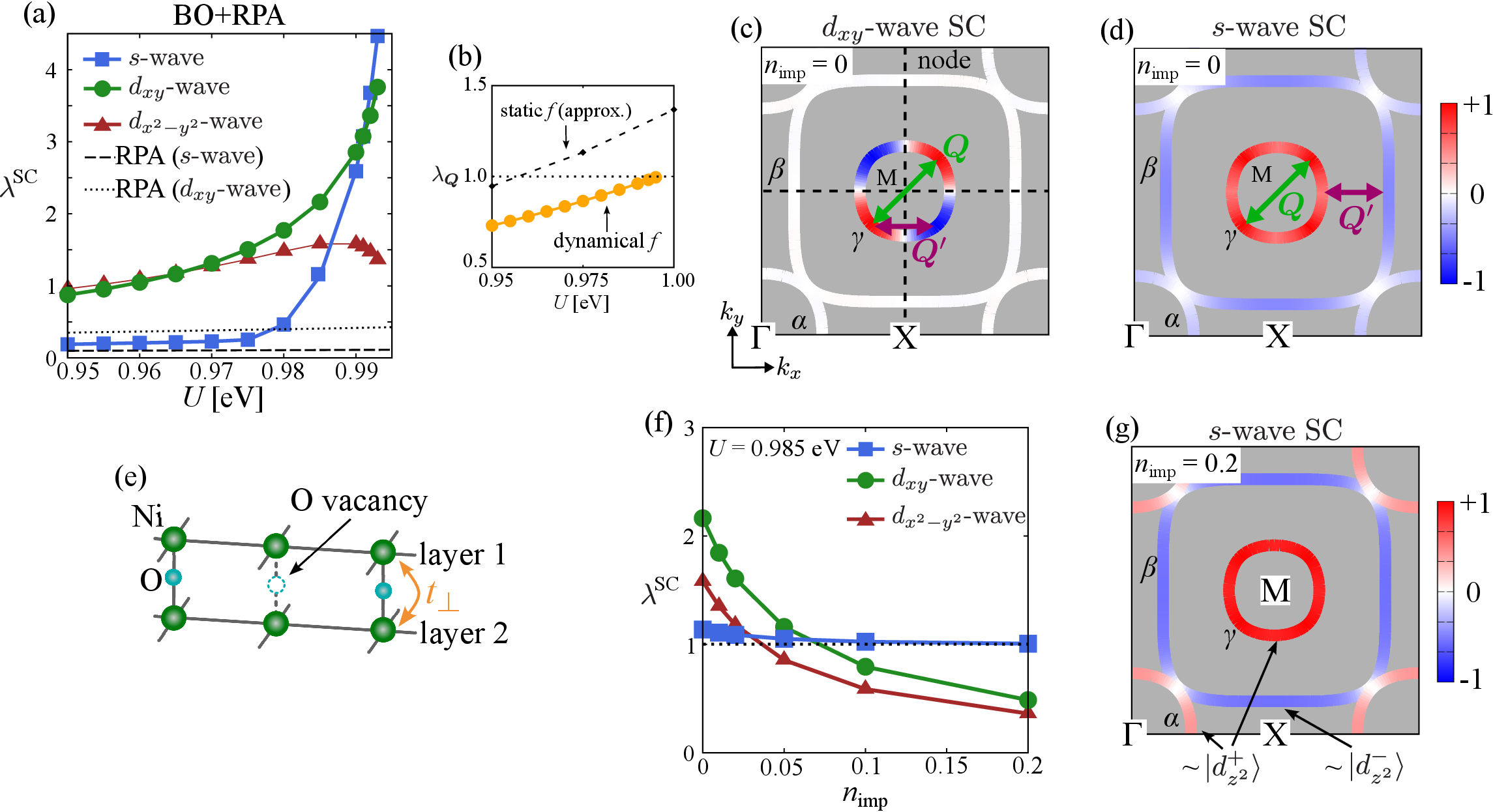}
	\caption{
{\bf Superconductivity: CDW + SDW fluctuation mechanism, effect of inner apical O vacancies}. \
(a) $U$-dependence of $\lambda^{\mathrm{SC}}$ derived from the linearized gap equation at $T=5$ meV. 
In the present mechanism, the SC eigenvalues of the $s$-wave state, $d_{xy}$-wave state, and $d_{x^2-y^2}$-wave state are magnified by the CDW fluctuations.
(In the RPA, in contrast, the obtained SC eigenvalues are small for the same $U$.)
(b) $U$-dependence of $\lambda_{\bm{Q}}$ derived from the DW equation. 
(c) Obtained $d_{xy}$-wave SC gap structure and (d) $s$-wave SC gap structure at $U=0.985$ eV, mediated by CDW + SDW fluctuations. 
Here, $\bm{Q}'$ and $\bm{Q}$ are the wavevectors for the maximum peaks of 
$\chi^{s}(\bm{q})$ and $\lambda_{\bm{q}}$, respectively, as shown in Fig. \ref{fig:fig3}. 
(e) Nonlocal impurity potential ${\hat O}_{\rm inner}$ due to inner apical O vacancy.
(f) Obtained $\lambda^{\mathrm{SC}}$ as functions of the O vacancy concentration $n_{\rm imp}$.
Only the $s$-wave SC state is robust against the inner apical O vacancies, so the relation $\lambda^{\mathrm{SC}}_{\rm s-wave}>\lambda^{\mathrm{SC}}_{\rm d_{xy}-wave}$ is realized for $n_{\rm imp}>0.06$.
(g) Obtained $s$-wave SC gap structure at $n_{\rm imp}=0.2$.
(The $s$-wave gap at $n_{\rm imp}=0$ is shown in (d).)
The sign-reversal between $\gamma$-pocket and $\beta$-pocket is robust, because ${\hat O}_{\rm inner}$ give no scattering between $|d_{z^2}^+\rangle$ and $|d_{z^2}^-\rangle$.
}
\label{fig:fig6}
\end{figure*}

Here, we study the origin of the high $T_{\mathrm{c}}$ superconductivity in La$_3$Ni$_2$O$_7$. 
We examine the SC pairing eigenvalue $\lambda^{\mathrm{SC}}$ and SC gap function based on the equation (\ref{eqn:gap}), by considering the CDW + SDW fluctuation-mediated pairing interaction $V^{\mathrm{SC}}=V^{\mathrm{bond}}-\frac{3}{2}{\hat C}^s{\hat\chi}^s(k-k'){\hat C}^s$:
See the Method section for further details.
We briefly explain the RPA results, where $V_{\mathrm{bond}}$ is negligibly small. 
In the RPA, both the $d_{xy}$-wave and $s_{\rm \pm}$-wave SC states are obtained, 
consistently with previous RPA analyses \cite{Eremin-Ni-SDW,Liu-Ni-SC2,Heier-Ni-SC}. 
As shown in Fig. \ref{fig:fig6} (a), the $\lambda^{\mathrm{SC}}$ within RPA is very small for $U\le1$ eV. (Note that it increases to $\sim2$ for $U=1.2$ eV, where $\alpha_s=0.94$.)

Next, we discuss the pairing mechanism due to the CDW + SDW fluctuation mechanism, by taking account of the VCs in ${\hat V}_{\mathrm{bond}}$. 
The obtained $\lambda^{\rm SC}$ of the $s$-wave, $d_{xy}$-wave, and $d_{x^2-y^2}$-wave SC states at $T=5$ meV are shown in Fig. \ref{fig:fig6} (a).
Importantly, all SC states are drastically stabilized by the CDW fluctuations.
The obtained $\lambda^{\mathrm{SC}}$ at $T=5$ meV exceeds unity thanks to large ${\hat V}_{\mathrm{bond}}$ although $\lambda^{\mathrm{SC}}$ is overestimated due to the absence of the self-energy. 
(Only $d_{x^2-y^2}$-wave SC state starts to saturate for $U\gtrsim0.99$ eV.)

Note that ${\hat V}_{\mathrm{bond}}(k,k')$ is strongly magnified for $\bm{k}-\bm{k}'\approx \bm{Q}$ in proportion to $(1-\lambda_{\bm{Q}})^{-1}$. 
Here, $\lambda_{\bm{Q}}$ is the eigenvalue of the DW equation derived from the ``dynamical $f$'' analysis. 
The obtained $\lambda_{\bm{Q}}$ is shown in Fig. \ref{fig:fig6} (b).

The derived SC gap functions on the FSs for the $d_{xy}$-wave and $s$-wave gap states are exhibited in Figs. \ref{fig:fig6} (c) and (d), respectively. 
In both gap states, the CDW fluctuations at $\bm{q}=\bm{Q}$ and the SDW fluctuations at $\bm{q}\approx\bm{Q}'$ enlarge $T_{\rm c}$ cooperatively, as denoted in Figs. \ref{fig:fig6} (c) and (d). 
Importantly, the $s$-wave gap in Fig. \ref{fig:fig6} (d) has sign-reversal between $\gamma$-pocket and $\alpha,\beta$-pockets with $|\Delta_\gamma|\gg|\Delta_{\alpha,\beta}|$. 
That is, ``prominent band-selective $s_{\pm}$-wave gap state'' is realized through the present CDW + SDW fluctuation pairing mechanism.

Note that the $\gamma$-pocket made of $d_{z^2}$-orbital is 
essential for the emergence of the bond-order fluctuations.
Importantly, the $\gamma$-pocket expands under applied pressure 
\cite{Sun-Ni-SC,Dagotto-Ni-SC2}
or through film thinning due to the self-doping
\cite{Ni-thin-CDMFT,Ni-thin-ARPES}.
Therefore, high-$T_{\rm c}$ superconductivity in La$_3$Ni$_2$O$_7$ can be induced either by applying pressure \cite{Sun-Ni-SC, Wang-Ni-SC, Zhang-Ni-SC} or by thinning the film \cite{Ko-Ni-SC, Zhou-Ni-SC}.
That will be caused by the combination of the CDW and SDW fluctuations. 
The fluctuations of vertical bond-orders shown in Fig. \ref{fig:fig4} (e) will mediate the strongest pairing glue. 
In SM F, we will explain $\bm{k}$, $\bm{k}'$-dependences of CDW and SDW fluctuations-mediated pairing interaction to discuss about the reason of the $d_{xy}$-wave and $s$-wave SC states are realized in more detail.

\section{$s$-wave SC state protected from oxygen vacancies}

Real samples of bilayer nickelate La$_3$Ni$_2$O$_{7-\delta}$ contains many oxygen vacancies; $\delta$ is about $0.1$ or even larger.
(In Ref. \cite{Wang-Ni-SC}, $T_{\rm c}\approx80$K is reported for $\delta=0.07$ under high pressure.)
Importantly, the oxygen vacancies primarily occupy the inner apical sites according to X-ray diffraction measurement \cite{vacancy}.
The inner O vacancies work as strong impurity potentials for the electron systems, because the large inter-layer hopping integral $t_\perp\ (=-0.66 \ {\rm eV}$ is killed by the O vacancy, as depicted in Fig. \ref{fig:fig6} (e).

As well-known, impurity effect is important to understand the mechanism and the symmetry of the pairing state.
In conventional isotropic $s$-wave SC state, $T_{\rm c}$ is robust against nonmagnetic impurities, known as Anderson theorem.
In highly contrast, anisotropic SC gap state with sign reversal, which is realized by spin-fluctuation mechanisms, is fragile against impurity scattering.
Importantly, the CDW fluctuations given by the present PMI mechanism mediates beyond-BCS strong attraction, leading to the heavily band-selective $s$-wave gap shown in Fig. \ref{fig:fig6} (d).
This PMI pairing mechanism has been successfully applied to understand various strongly-correlated superconductors close to the CDW instabilities. 
The $s$-wave SC states predicted by this mechanism is robust against impurities, as observed in Fe-based superconductors \cite{Prozorov-Fe, Ghigo-Fe}, kagome metals \cite{Roppongi-kagome-imp}, and the heavy fermion compound CeCu$_2$Si$_2$ \cite{Yamashita-heavy-fermion}.

In general, the impurity effect on $T_{\rm c}$ for a given pairing state will depend on the major impurity sites.
Fortunately, we know that the major impurity scattering in La$_3$Ni$_2$O$_{7-\delta}$ is caused by the inner apical oxygen vacancy shown in Fig. \ref{fig:fig6} (e).
In the following, we demonstrate that the $s$-wave SC state in Fig. \ref{fig:fig6} (d) is robustly protected from inner apical oxygen vacancies.
The present study provide valuable insights into the pairing mechanism of bilayer nickelates.


In the real-space $d_{z^2}$-orbital basis $\{ |d_{z^2}^{1}\rangle, |d_{z^2}^{2}\rangle \}$, the impurity potential in Fig. \ref{fig:fig6} (e) is expressed as ${\hat O}^{\rm inner}=w{\hat \tau}_x$, where $\tau_\mu$ is the Pauli matrix.
It is converted to $w{\hat \tau'}_z$ in the bonding and antibonding $d_{z^2}$-orbital basis $\{ |d_{z^2}^{+}\rangle, |d_{z^2}^{-}\rangle \}$, where $|d_{z^2}^{\pm}\rangle=(|d_{z^2}^{1}\rangle \pm |d_{z^2}^{2}\rangle)/\sqrt{2}$.
Therefore, ${\hat O}^{\rm inner}$ becomes diagonal with respect to the $d_{z^2}^\pm$-orbital basis.
Below, we set $w=-t_\perp$ to cancel the original inter-layer hopping integral.
(Note that a different impurity potential $\hat{O}^{\rm inner}=w'{\hat \tau}_0$ is also converted to the diagonal form $w'{\hat \tau'}_0$ in the $d_{z^2}^\pm$-orbital basis.)

We analyze the impurity effect based on the $T$-matrix due to the single impurity:
${\hat T}(\e_n)= {\hat O}^{\rm inner}(1-{\hat O}^{\rm inner}\hat{g}_{\rm loc}(\e_n))^{-1}$,
where $\hat{g}_{\rm loc}(\e_n)=\frac1N \sum_\k {\hat G}(\k,\e_n)$ is the local Green function.
Using the $T$-matrix, the impurity-induced normal and anomalous self-energies are ${\hat \Sigma}(\e_n)=n_{\rm imp}{\hat T}(\e_n)$ and ${\hat \Delta}(\e_n)=n_{\rm imp}|{\hat T}(\e_n)|^2 f_{\rm loc}(\e_n)$, respectively.
Here, $n_{\rm imp} \ (=\delta)$ is the concentration of inner apical O vacancy, and $f_{\rm loc}(\e_n)=\frac1N \sum_\k {\hat G}(k){\hat \Delta}(k) ^t{\hat G}(-k) + O(\Delta^3)$.
Based on the $T$-matrix theory, we can extend the SC gap equation in the presnce of dilute impurities; see Ref. \cite{Onari-imp} for detailed explanations.

Figure \ref{fig:fig6} (f) exhibits the obtained eigenvalues for the $s$-wave and $d$-wave SC states as functions of $n_{\rm imp}$ at $U=0.985$ eV.
Interestingly, the $s$-wave SC state is robust against the inner apical O vacancies, so the relation $\lambda^{\mathrm{SC}}_{\rm s-wave}>\lambda^{\mathrm{SC}}_{\rm d_{xy}-wave}$ is realized for $n_{\rm imp}>0.06$.
The $s$-wave gap function at $n_{\rm imp}=0.2$ is shown in Fig. \ref{fig:fig6} (g).
By impurity scatterings,
finite SC gap is induced in the $\a$-pocket, while the sign reversal between $\gamma$- and $\beta$-pockets at $n_{\rm imp}=0$ in Fig. \ref{fig:fig6} (d) is intact against large impurity scattering.
The reason is that the $\g$-pocket is mainly composed of the $d_{z^2}^+$-orbital, and the $\a$-pocket contains the $d_{z^2}^+$-orbital component.
Because the $\beta$-pocket contains the $d_{z^2}^-$-orbital component, impurity scattering between $\gamma$- and $\beta$-pockets are almost absent.
This study strongly indicates the realization of the $s$-wave SC state in bilayer nickelates, considering its robustness against the inner apical O vacancies \cite{vacancy} as we show in Fig. \ref{fig:fig6} (f).


Finally, we emphasize that Ni-site impurities lead to significant scattering between all $\alpha$-, $\beta$-, and $\gamma$-pockets \cite{Eremin-imp}. As a result, for the $s$-wave SC state, Ni-site impurities cause a stronger suppression compared to the inner apical oxygen vacancies. Moreover, due to the strong CDW fluctuations, Ni-site impurities promote the $s$-wave SC state without sign reversal, as observed in kagome metals \cite{Tazai-kagome1,Roppongi-kagome-imp} and Fe-based SCs \cite{Onari-imp,Prozorov-Fe,Ghigo-SC}. Therefore, investigating the effects of impurities will provide valuable insights into the pairing mechanism.


\section{Discussions}

\subsection{Origin of CDW formation, CDW + SDW fluctuation-mediated SC}

The aim of this paper is to uncover the origin of the CDW + SDW instability in nickelates, which is widely observed but remains unresolved.
To address this issue,
we demonstrate that both inter- and intra-layer CDW instabilities develop due to the PMI mechanism, which captures important electron correlations beyond mean-field theory \cite{Kontani-rev2021}. 
Notably, $d_{z^2}$-orbital vertical bond-order instability develops concurrently with the SDW instability. 
The present theory naturally explains why the CDW instability in bilayer nickelates is comparable to, or even larger than, the SDW instability.
Furthermore, the experimental double-stripe CDW/SDW state is naturally explained.
We find that the coexistence of charge and spin fluctuations is essential in bilayer nickelates, with both playing a cooperative role in mediating high-$T_c$ superconductivity.

In this mechanism, the CDW instability is very sensitive to the size of the $d_{z^2}$-orbital hole pocket, which can be efficiently controlled by applied pressure \cite{Sun-Ni-SC, Wang-Ni-SC, Zhang-Ni-SC} or film thinning \cite{Ko-Ni-SC, Zhou-Ni-SC}.
In the SI B,
we discuss the effect of carrier-doping on the CDW formation. 
Due to the hole-doping ($n<3.0$),
strong CDW instability survives although its strength is moderately decreased while ${\bm Q}$ slightly changes. 
Thus, the CDW order by the PMI mechanism is relatively robust against the hole-doping. 
In contrast, due to the electron-doping ($n>3.0$), the CDW instability is suddenly reduced as $\gamma$-pocket shrinks, indicating the significance of the $\gamma$-pocket in the PMI mechanism. 
This result would explain why $T_{\rm cdw}$ sensitively depends on the sample quality in La$_3$Ni$_{2}$O$_{7}$.

\subsection{Comment on high-$T_c$ in bilayer nickelate thin films at ambient pressure}

In the present theory, the obtained CDW + SDW fluctuations cooperatively mediate $s$-wave and $d$-wave high-$T_{\rm c}$ superconductivity in nickelates. 
Very recently, superconductivity at ambient pressure ($T_{\rm c}\approx 40$ K) has been discovered in bilayer nickelate thin films \cite{Ko-Ni-SC,Zhou-Ni-SC}.
Importantly, electron filling is $n\approx2.6$ \cite{Ni-thin-CDMFT,Ni-thin-ARPES}, while the size of the $d_{z^2}$-orbital hole-pocket observed by ARPES measurement is comparable to that in the present model (shown in Fig. \ref{fig:fig1} (c)) \cite{Ni-thin-ARPES}.
Therefore, the current theory, which highlights the significance of the $d_{z^2}$-orbital hole pocket, offers crucial insights into the investigation of DW orders in nickelate thin films, which remain unexplored.

In bilayer nickelate thin films \cite{Ko-Ni-SC,Zhou-Ni-SC}, the reported onset $T_{\rm c}$ reaches 40 K, while the residual resistivity is relatively large (about 0.2 m$\Omega \cdot$cm.), suggesting that the SC state is robust against impurities. 
This fact strongly indicates the realization of the $s$-wave SC state, which is robust against the inner apical O vacancies \cite{vacancy} as shown in Fig. \ref{fig:fig6} (f).
The $s$-wave state with sign reversal, shown in Fig. \ref{fig:fig6} (d), will be suppressed by Ni-site impurities, although it remains more robust than the nodal $d$-wave states.
Note that the impurity effects on $T_{\rm c}$ have been extensively studied in Fe-based superconductors \cite{Prozorov-Fe, Ghigo-Fe}, kagome metals \cite{Roppongi-kagome-imp}, and the heavy fermion compound CeCu$_2$Si$_2$ \cite{Yamashita-heavy-fermion}, to achieve valuable insights into the pairing mechanism. 


\section{Methods}



\subsection{Density-wave equation analysis}

In order to explore the CDW order, we have to investigate beyond-RPA nonlocal electron correlations, described as the VCs.
The DW equation method enables us to introduce the AL-type and MT-type VCs iteratively, which are necessary to preserve the conserving laws.
The derived quantum phase transition from the DW equation automatically minimizes the Luttinger-Ward free energy in Fermi liquids \cite{Tazai-LW}.

\begin{figure}[!htb]
	\includegraphics[width=.99\linewidth]{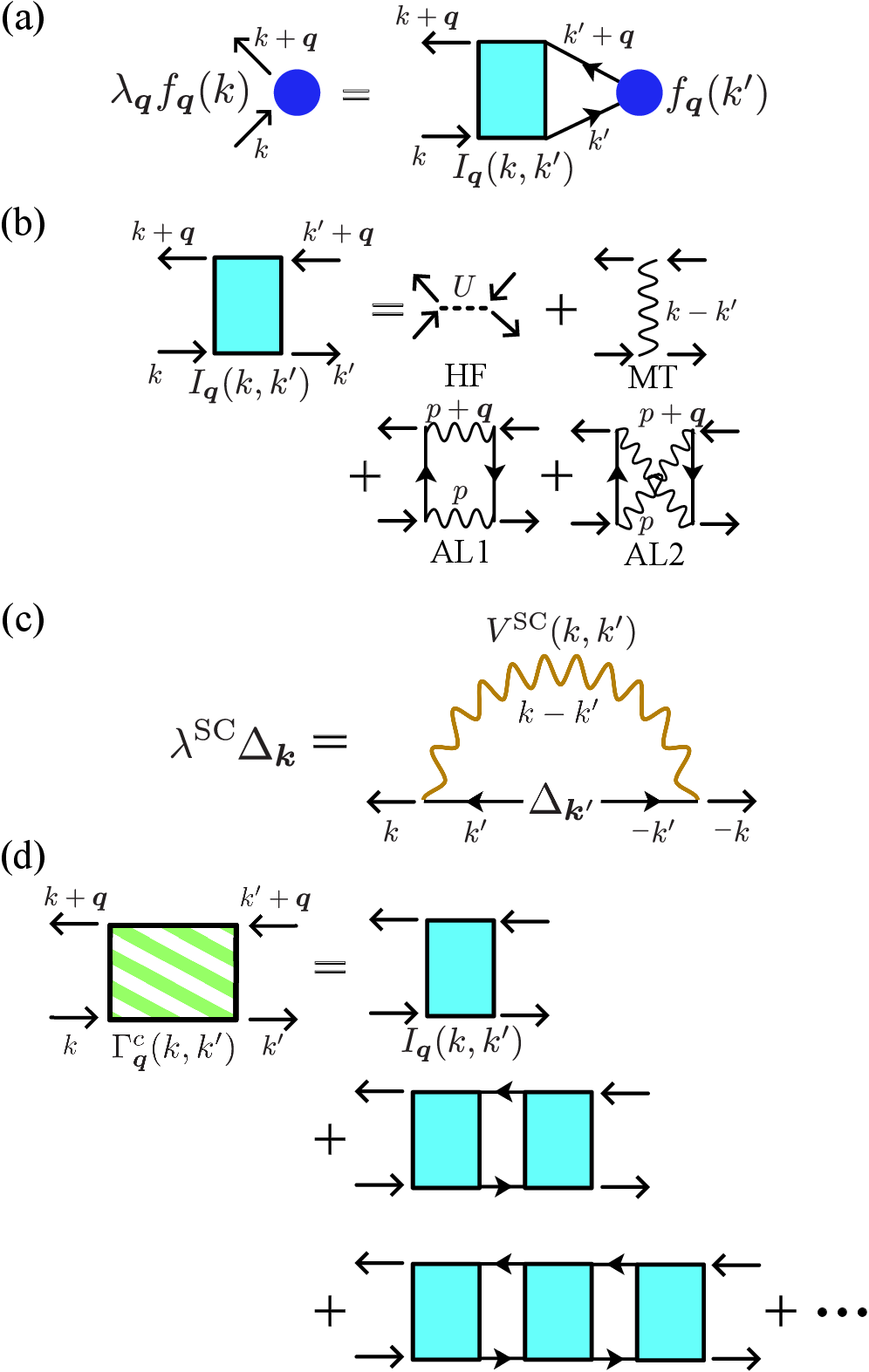}
	\caption{
{\bf Diagrammatic expressions of DW equation and Bethe-Salpeter equation}. \
(a) Diagrammatic expression of the linearized DW equation.
$f_{\bm{q}}(k)$ is the charge-channel form factor and $\lambda_\q$ is the eigenvalue.
(b) Charge-channel kernel function $I_{\bm{q}}(k,k')$ ({\it i.e.}, the irreducible four-point vertex) composed of Hartree term,
MT term, and two AL terms.
Each wavy line represents spin and charge fluctuations-mediated interaction.
(c) Linearized SC gap equation.
The yellow wavy line represents the effective pairing interaction derived from the Bethe-Salpeter equation. 
(d) Diagrammatic expression of the charge-channel full four-point vertex $\Gamma^c_{\bm{q}}(k,k')$ derived from the Bethe-Salpeter equation.
$\Gamma_{\bm{q}}^c$ diverges when $\lambda_{\bm{q}}$ reaches unity.
}
\label{fig:fig2}
\end{figure}

Here, we introduce the linearized DW equation for the charge-channel order,
which is schematically shown in Fig. \ref{fig:fig2} (a).
Its analytic expression is 
\begin{eqnarray}
	\begin{split}
		\lambda_{\bm{q}}f_{\bm{q}}^{L}(k) &= -\frac{T}{N}\sum_{k',M_1,M_2}I_{\bm{q}}^{L,M_1}(k,k') \\
		&\times \{\hat{G}(k')\hat{G}(k'+\bm{q})\}^{M_1,M_2} f_{\bm{q}}^{M_2}(k'),
		\label{eqn:DW}
	\end{split}
\end{eqnarray}
where $k\equiv (\bm{k},\epsilon_n)$, and $\epsilon_n=(2n+1)\pi T$.
$L \equiv (l,l')$ and $M_i \equiv (m_i,m_i')$ represent the pair of layer indices.
Here, the eigenfunction $f_{\bm{q}}^{L}(k)$ is called the form factor, which represents the non-local charge-channel DW order.
The eigenvalue $\lambda_{\bm{q}}$ reaches unity when the DW order occurs at wavevector $\bm{q}$.
In the DW equation scheme, the DW is the ``electron-hole pairing'' mediated by the kernel function ${\hat I}_{\bm{q}}(k,k')$, which determines the accuracy of the approximation.
To satisfy the thermal equilibrium condition,
the kernel function should satisfy the Ward identity \cite{Tazai-LW}:
$\hat{I}_{\bm{q}}$ is given by the Fourier transformation of 
$I_{ij,i'j'}\equiv\frac{\delta^2 \Phi_{\mathrm{LW}}}{\delta G_{ij}\delta G_{i'j'}}$, 
where $\Phi_{\mathrm{LW}}$ is the Luttinger-Ward function, and $ij,i'j'$ are site indices. 
For $\bm{q}=\bm{0}$, it is simply given as $I^{LM}_{\bm{q}=\bm{0}}(k,k')=\delta^2 \Phi_{\mathrm{LW}}/\delta G_L(k) \delta G_M(k')$.
By applying the one-loop approximation for $\Phi_{\mathrm{LW}}$,
the derived kernel function is given in Fig. \ref{fig:fig2} (b) schematically, 
and its analytic expression is
\begin{eqnarray}
&&\hat{I}=\hat{I}_{\bm{q}}^{\mathrm{Hartree}}+\hat{I}_{\bm{q}}^{\mathrm{MT}}+\hat{I}_{\bm{q}}^{\mathrm{AL1}}+\hat{I}_{\bm{q}}^{\mathrm{AL2}}, \label{I_tot}\\
&&I_{\bm{q}}^{\mathrm{Hartree};ll',mm'}(k,k')=C^c_{ll',mm'}, \label{I_hartree}\\
&&I_{\bm{q}}^{\mathrm{MT};ll',mm'}(k,k')=-\frac{a^b}{2}\sum_{b=s,c} \left[V^b_{lm,l'm'} (k-k') \right. \notag \\
&&\ \ \ \ \ \ \ \ \ \ \ \ \ \ \ \ \ \ \ \ \ \ \ \ \ \ \ \ \ \ \ \ \ \ \ \ \ \ \ \ \ \ \ \ \left. -C^b_{ll',mm'}\right],  \label{I_MT} \\
&&I_{\bm{q}}^{\mathrm{AL1};ll',mm'}(k,k') \notag \\
&&=\frac{T}{N}\sum_{b=s,c}\sum_{\substack{p,l_1l_2 \\ m_1m_2}}\frac{a^b}{2} V^b_{l l_1,m m_1}(p+\bm{q})V^b_{m' m_2,l' l_2}(p) \notag \\
&&\ \ \ \ \ \ \ \ \ \ \ \ \ \ \ \ \ \ \ \ \times G_{l_1l_2}(k-p)G_{m_2m_1}(k'-p), \label{I_AL1}\\
&&I_{\bm{q}}^{\mathrm{AL2};ll',mm'}(k,k') \notag \\
&&=\frac{T}{N}\sum_{b=s,c}\sum_{\substack{p,l_1l_2 \\ m_1m_2}}\frac{a^b}{2} V^b_{l l_1,m_2 m'}(p+\bm{q})V^b_{m_1 m,l' l_2}(p) \notag \\
&&\ \ \ \ \ \ \ \ \ \ \ \ \ \ \ \ \ \ \ \ \times G_{l_1l_2}(k-p)G_{m_2m_1}(k'+p+\bm{q}), \label{I_AL2}
\end{eqnarray}
where $p\equiv (\bm{p},i\omega_l)$ and $b=c \ (s)$ is the index of the charge (spin) channel.
Here, $a^{s}=3$, $a^{c}=1$, and
$\hat{V}^{b}\equiv \hat{C}^b+\hat{C}^b\hat{\chi}^b\hat{C}^b$ is the $b$-channel interaction matrix.
Note that the double-counting in the $U^2$-terms in Eqs. (\ref{I_hartree})-(\ref{I_AL2}) should be subtracted properly. 
The first-order term with respect to $U$ in $\hat{I}$ (Hartree-term) is given by 
Eq. (\ref{I_hartree}).
The present bond-order is mainly derived from the two AL terms expressed in Eqs. (\ref{I_AL1}) and (\ref{I_AL2}), while the contribution from the Hartree-term is almost zero.

\subsection{Beyond-Migdal superconducting gap equation}
We also study the unconventional superconductivity mediated by 
CDW fluctuations, 
which cause attractive pairing interaction 
\cite{Tazai-heavy-fermion3,Kontani-lc}.
We solve the following linearized SC gap equation on the FSs:
\begin{eqnarray}
\lambda^{\mathrm{SC}} \Delta_{\bm{k}}(\epsilon_n) = \frac{\pi T}{(2\pi)^2} \sum_{\epsilon_{n'}} \oint_{\mathrm{FSs}}
\frac{d\bm{k}'}{v_{\bm{k}'}} \frac{\Delta_{\bm{k}'}(\epsilon_{n'})}{|\epsilon_{n'}|} V^{\mathrm{SC}}(k,k'),
\label{eqn:gap}
\end{eqnarray}
where $v_{\bm{k}}$ is the Fermi velocity. 
Its diagrammatic expression is given in Fig. \ref{fig:fig2} (c).
Here, $\Delta_{\bm{k}}(\epsilon_n)$ is the gap function on the FSs,
and the eigenvalue $\lambda^{\mathrm{SC}}$ reaches unity at $T=T_{\rm c}$.
$V^{\mathrm{SC}}(k,k')$ in Eq. (\ref{eqn:gap}) 
is the singlet pairing interaction mediated by CDW and SDW fluctuations 
in the band representation.
It is derived from the following layer (or orbital) representation function obtained by the Green function method:
\begin{eqnarray}
{\hat V}^{\mathrm{SC}}(k,k') &=& {\hat V}_{\mathrm{bond}}(k,k')
\nonumber \\
& & -\frac{3}{2}{\hat C}^s{\hat\chi}^s(k-k'){\hat C}^s - \frac12 ({\hat C}^s-{\hat C}^c) ,
	\label{eqn:V_SC}
\end{eqnarray}
which is converted to the band representation by using the matrix element of the unitary transformation from layer $l$ to band $b$: $U_{l,b}(\k)\equiv\langle l,k|b,\k\rangle$.

%

The first term of Eq. (\ref{eqn:V_SC})
represents the attractive interaction due to bond-order fluctuations.
It is expressed as $V_{\rm bond}\approx f_{\bm{k}-\bm{k}'}(\bm{k}')f_{\bm{k}'-\bm{k}}(-\bm{k}') \frac{g_{\bm{k}-\bm{k}'}}{1-\lambda_{\bm{k}-\bm{k}'}}$ based on the DW equation theory
\cite{Tazai-heavy-fermion3,Kontani-lc}. 
Here, the function $g_\q$ is given as $\frac1{N^2} \sum_{k,k'} f_\q(k) f_{-\q}(k'+\q)I_\q(k,k')/\frac1{N^2} \sum_{k.k'} |f_\q(k)|^2 |f_{-\q}(k+\q)|^2$ by using the kernel function of the DW equation.
The magnitude of $g_\q$ can exceed $U$ in kagome lattice Hubbard models \cite{Tazai-kagome1}.
Unfortunately, it is difficult to obtain the $\bm{k}$-dependent functions $g_\k$ and $\lambda_\k$ accurately.
To resolve this problem, we perform the numerical analysis of the Bethe-Salpeter (BS) equation to obtain the function of $V_{\rm bond}(k,k')$ accurately for the Fermi momenta $\k$ and $\k'$.
The BS equation for the charge-channel full four-point vertex $\hat\Gamma^c_{\bm{q}}$ is given as 
\begin{eqnarray}
{\hat\Gamma}^c_{\bm{q}}(k,k')&=&{\hat I}_{\bm{q}}(k,k') \notag \\
 &&- \frac{T}{N}\sum_{p} {\hat I}_{\bm{q}}(k,p) {\hat G}(p){\hat G}(p+\bm{q}) {\hat \Gamma}^c_{\bm{q}}(p,k'),
\label{eqn:BSeq}
\end{eqnarray}
where $\hat I$ is the kernel function of the DW equation.
Its diagrammatic expression is given in Fig. \ref{fig:fig2} (d).
Then, the charge-channel pairing interaction, 
which is the first term of Eq. (\ref{eqn:V_SC}), is given as
$V_{\rm bond}({\bm k},{\bm k}')=\frac12 {\hat\Gamma}^c_{{\bm k}'-\bm{k}}({\bm k},-{\bm k}')-\frac{1}{2}{\hat C}^c$.
(In $V_{\mathrm{bond}}$, the $U$-linear term is dropped.)
Note that the RPA result $-\frac{U^2}{2} \chi^c(k-k')$ is reproduced by considerint only the Hartree term $U$ in $I$ from Eq. (\ref{eqn:V_SC}).
In the RPA, $V_{\rm bond}$ is negative and small in magnitude.
However, $V_{\rm bond}$ becomes positive and strongly magnified due to VCs for the PMI processes (shown in Fig. \ref{fig:fig2} (b)), by reflecting the strong bond-order fluctuations obtained by the DW equation analysis.
Based on the BS equation method, $V_{\rm bond}(k,k')$ is derived without numerical ambiguity.
$V_{\mathrm{bond}}$ given by Eq. (\ref{eqn:BSeq}) diverges when $\lambda_{\bm{q}}=1$. 
The BS equation method was previously used to 
derive the bond-order-fluctuation mediated pairing interaction 
in kagome metal by focusing on only three van-Hove singularity points 
\cite{Tazai-kagome1}.
Here, we solve Eq. (\ref{eqn:BSeq}) for the Fermi momenta $\bm{k}$ and $\bm{k}'$, divided the FSs into 48 patches. 
The detailed explanation of the BS equation method will be presented in Ref. \cite{Yamakawa-BS}.









\clearpage


\makeatletter
\renewcommand{\thefigure}{S\arabic{figure}}
\renewcommand{\theequation}{S\arabic{equation}}
\makeatother
\setcounter{figure}{0}
\setcounter{equation}{0}
\setcounter{page}{1}
\setcounter{section}{1}

\begin{center}
{\bf \large 
[Supplementary information] \\ 
\vspace{3mm}
{\large
Unified mechanism of charge-density-wave and high-$T_{\rm c}$ superconductivity protected from oxygen vacancies in bilayer nickelates
}
}
\end{center}

\begin{center}
Daisuke Inoue$^{1}$, Youichi Yamakawa$^{1}$, Seiichiro Onari$^{1}$, and Hiroshi Kontani$^{1}$
\end{center}

\begin{center}
\textit{
$^1$Department of Physics, Nagoya University, Nagoya 464-8602, Japan 
}
\end{center}


\section*{A: RPA and DW analysis based on the two-orbital $H_U$ model}

\begin{figure}[!htb]
	\includegraphics[width=.99\linewidth]{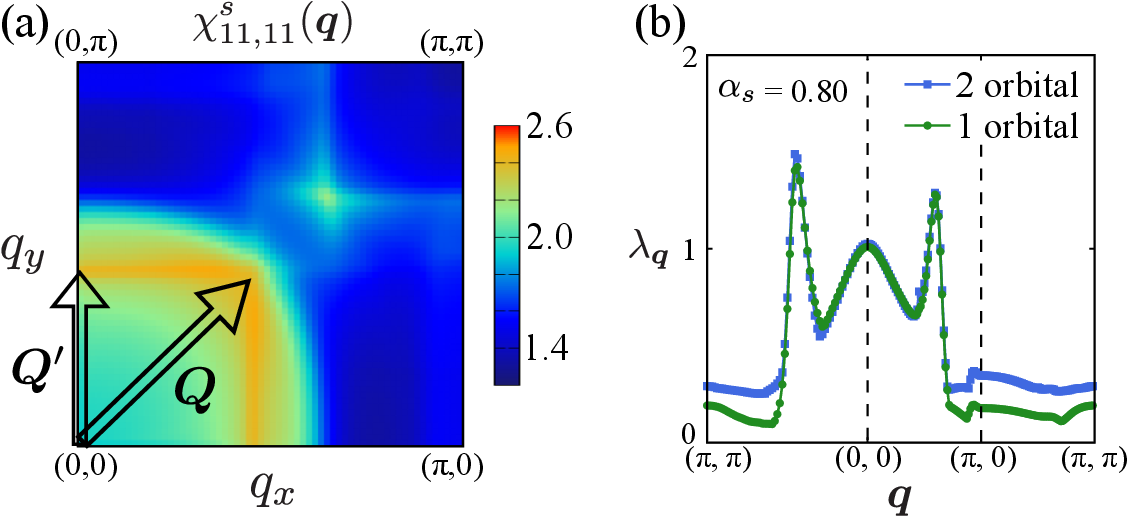}
	\caption{
(a) $\bm{q}$-dependences of the RPA $\chi^s_{11,11}(\bm{q})$ based on 
the two-orbital ($d_{z^2}$- and $d_{x^2-y^2}$-orbital) Coulomb interaction model. 
Importantly, the obtained $\chi^s$ shows maximum peak at $\bm{q}\approx\bm{Q}$.
(b) Comparison between $\lambda_{\bm{q}}$ given by the single-orbital $H_U$ model and two-orbital $H_U$ model for $\alpha_s=0.80$. 
Both results are quantitatively equivalent.
(Here, $U=1.02$ eV ($1.03$ eV) for the single-orbital (two-orbital) $H_U$ model.)
Note that (a)-(b) are derived from the ``static $f$'' analysis of the DW equation.
	}
\label{fig:figA1}
\end{figure}

Here, we calculate the RPA spin susceptibility, 
$\chi_{l_1 l_2, l_3 l_4}^{s, \ m_1 m_2, m_3 m_4}(\q)$, in the present La$_3$Ni$_2$O$_7$ tight-binding model 
by considering the two-orbital Coulomb interaction ${\hat C}^b$ ($J=U/10$, $U'=U-2J$) for $d_{x^2-y^2}$- and $d_{z^2}$-orbitals. 
The expression of multi-orbital ${\hat C}^b$ is presented in Ref. \cite{SKontani-rev2021}.
Here, $l_n \ (=1,2)$ is the layer index, and $m_n\ (=d_{z^2}, d_{x^2-y^2})$ is the orbital index. 
Figure \ref{fig:figA1} (a) shows obtained $\bm{q}$-dependences of the intra-layer RPA spin susceptibility $\chi_{11,11}^{s}(\q)$ 
based on the two-orbital ($d_{z^2}, d_{x^2-y^2}$) Coulomb interaction ($H_U$) model, 
where $\chi_{l_1 l_2, l_3 l_4}^{s}(\q)\equiv \sum_{m_1,\cdots, m_4} \chi_{l_1 l_2, l_3 l_4}^{s, \ m_1 m_2, m_3 m_4}(\q)$. 
The obtained $\chi^s_{11,11}$ takes maximum value around $\q\approx\bm{Q}$.
Note that the relation $\chi^s_{11,11}(\bm{Q})\gtrsim\chi^s_{11,11}(\bm{Q}')$ is clearly realized compared with the results of the single-orbital model given in the main text. 



Also, we perform the DW equation analysis 
based on the single $d_{z^2}$-orbital $H_U$ model and 
the two-orbital ($d_{z^2}, d_{x^2-y^2}$) Coulomb interaction $H_U$ model. 
(The expression of the DW equation for multi-orbital $H_U$ models is presented in Ref. \cite{SKontani-rev2021}.)
Figure \ref{fig:figA1} (b) shows the obtained eigenvalues for $\alpha_s=0.8$, where $U=1.02$ eV ($1.03$ eV) for the single-orbital (two-orbital) $H_U$ model.
It is verified that obtained $\lambda_{\bm{q}}$ by the two-orbital DW equation 
is very similar to that by the single-orbital calculation. 
Thus, the numerical results based on the single $d_{z^2}$-orbital $H_U$ model presented in the main text are reliable.

\section*{B: CDW and SDW instabilities for hole- and electron-doping}


\begin{figure}[!htb]
	\includegraphics[width=.99\linewidth]{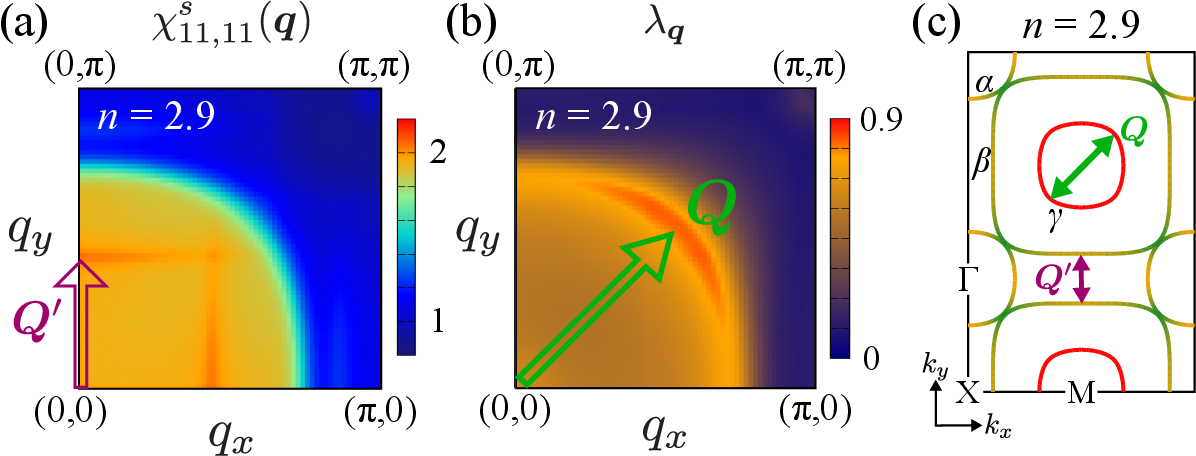}
	\caption{
(a)(b) Obtained $\bm{q}$-dependences of (a) $\chi^s_{11,11}(\bm{q})$ 
and (b) $\lambda_{\bm{q}}$ at $n=2.9$. 
Note that (b) are derived from the ``dynamical $f$'' analysis of the DW equation. 
(c) Main nesting vector $\bm{Q}$ and $\bm{Q}'$ at $n=2.9$:
$\bm{Q}$ is given by the intra $\gamma$-pocket nesting, 
which corresponds to the maximum peak of $\lambda_{\bm{q}}$ shown in (b).
$\bm{Q}'$ is given by the inter ($\beta$-$\beta$) pocket nesting,
which corresponds to the peak of $\chi^s(\q)$ shown in (a).
}
\label{fig:figA4}
\end{figure}


We discuss the hole-doping case ($n<3.0$).
In this case, $\lambda_{\bm{Q}}$ moderately decreases, as we show in Fig. \ref{fig:fig3} (g). 
The obtained $\chi^{s}_{11,11}(\bm{q})$ and $\lambda_{\bm{q}}$ at $n=2.9$
are shown in Figs. \ref{fig:figA4} (a) and (b), respectively.
Here, $U=1.0$ eV and $\alpha_s=0.75$. 
We find that $\chi^s(\bm{q})$ and $\lambda_{\bm{q}}$ 
takes maximum value around $\bm{q}\approx \bm{Q}'$ and 
$\q\approx \bm{Q}$ originated from the Fermi surface nesting,
similarly to $n=3.0$ case.
The nesting vectors at $n=2.9$ are depicted in Fig. \ref{fig:figA4} (c).
Therefore, the CDW order due to the PMI mechanism is robust even for 
the La$_3$Ni$_2$O$_7$ models with larger $\gamma$-pocket.
Then, $|\bm{Q}|$ is slightly enlarged by hole-doping 
because the $\gamma$-pocket is magnified slightly.
The schematic picture of the inter/intra-layer bond-order is shown in Figs. \ref{fig:fig4} (e) and (f) in the main text.



\section*{C: Single-$\q$ CDW state, Coexistence of single-${\bm Q}_1$ CDW and single-${\bm Q}_2$ SDW}

\begin{figure}[!htb]
\includegraphics[width=.99\linewidth]{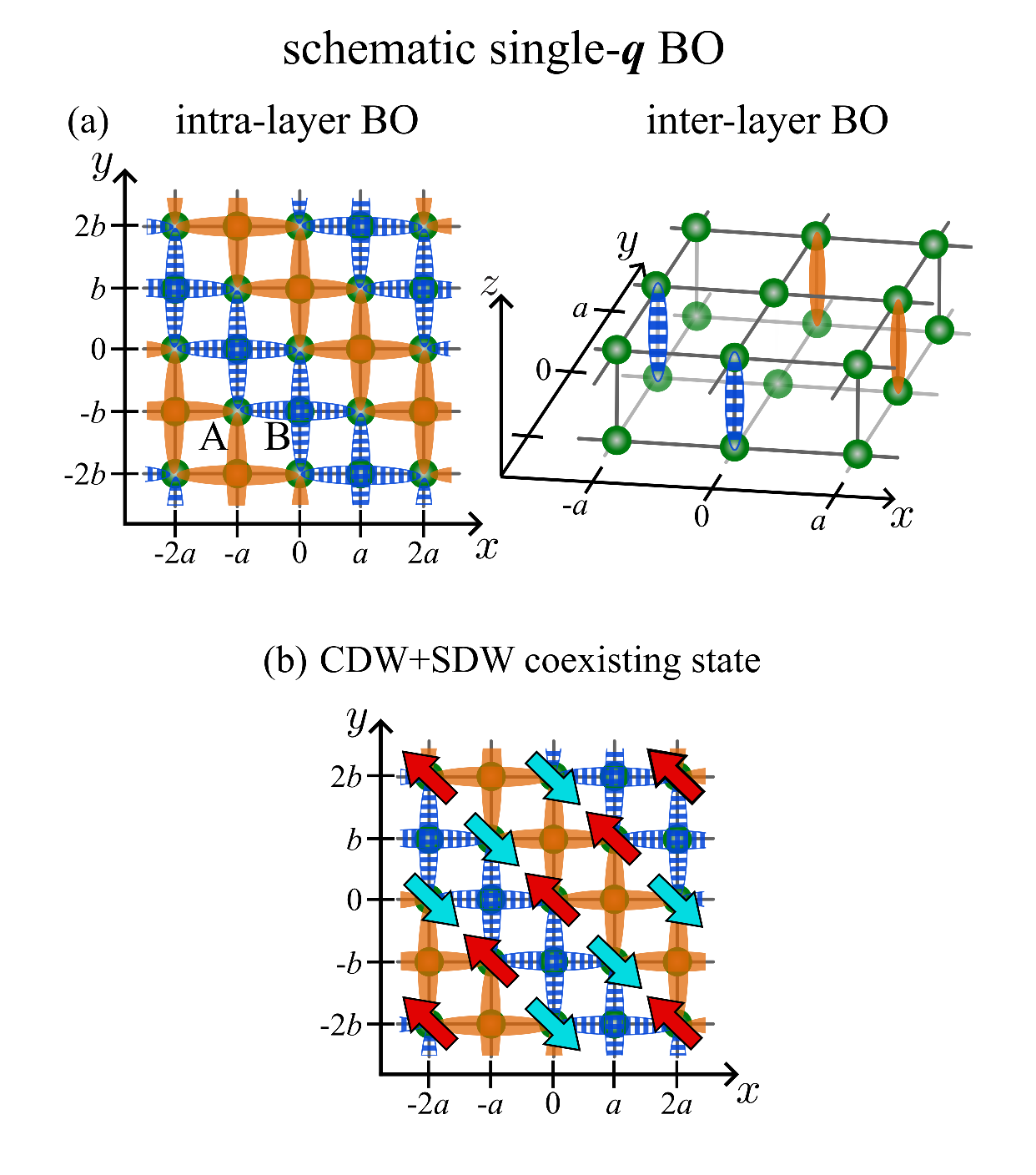}
\caption{
(a) Simple schematic pictures of a possible single-$\bm{q}$ inter/intra-layer bond-order state at wavevector $\q_{\rm c}=\bm{Q}_1$. 
An orange-filled (blue-striped) bond represents the hopping integrals modulated by $+\delta t >0$ ($-\delta t <0$). 
Here, we denote only the vertical and the third-nearest bond modulations to simplify the picture.
(b) Possible SDW at wavevector $\q_{\rm s}=\bm{Q}_2$ in the bond-order phase.
The red (light blue) arrows denote the spin-up (down) sites.
We denote only the sites with large spin moments, as indicated by the results of $\mu$SR and neutron diffraction measurements \cite{Sspin-moment}.
}
\label{fig:figA6}
\end{figure}

Figure \ref{fig:figA6} (a) exhibits a simplified picture of single-$\bm{q}$ inter/intra-layer bond-order state in real space, derived from the form factor $\tilde{F}^{ll'}(\bm{r})$ shown in Figs. \ref{fig:fig4} (e) and (f).
The schematic picture is derived from $\delta t_{ij}$ given as Eq. (\ref{eqn:hopping_modulation}) with $\bm{q}=\bm{Q}_1$, by setting $\psi=\pi/2$. 
Here, $(x,y)$ is the intra-layer Ni site.
Orange-filled (blue-striped) bond represents the 
hopping integrals modulated by $+\delta t >0$ ($-\delta t <0$). 
We stress that the largest bond-order parameter is the vertical inter-layer hopping modulation. 

Under the single-$\bm{q}$ bond-order, the four-period hopping modulation causes
local inversion symmetry breaking at site A in Fig. \ref{fig:figA6} (a). 
On the other hand, the global inversion symmetry is not broken, as understood by focusing on site B.
Note that the local inversion symmetry breaking is suggested by the recent X-ray measurement \cite{SRen-Ni-X-ray}.

Next, we show a possible CDW + SDW coexisting state with $\q_{\rm s}\approx\bm{Q}_2$ in Fig. \ref{fig:figA6} (b). 
Here, the red (light blue) arrows indicate the spin-up (down) site on the ab-plane. 
In general, the CDW at $\bm{q}_{\rm c}={\bm Q}$ and the SDW at $\bm{q}_{\rm s}={\bm Q}'$ easily coexist because they use different nestings of the FS. 
The CDW + SDW coexisting state in Fig. \ref{fig:figA6} (b) might be consistent with previous experimental reports such as X-ray measurements
\cite{SXie-Ni-neutron,SLiu-Ni-resistivity,SWu-Ni-DW,SMukuda-Ni-NMR,SChen-Ni-mu+SR,SKhasanov-Ni-muSR,SChen-Ni-RIXS,SRen-Ni-X-ray}. 
In Fig. \ref{fig:figA6} (b), we denote only the sites with large spin moments, as indicated by the results of $\mu$SR and neutron diffraction measurements \cite{Sspin-moment}.

\section*{D: Double-$\bm{q}$ CDW state, Coexistence of double-$\q$ CDW and double-$\q$ SDW}

In this paper, we obtain the inter/intra-layer bond-order with $\bm{q}=\bm{Q}$ based on the present DW equation analysis. 
Both the "single-$\bm{q}$ bond-order" and "double-$\bm{q}$ bond-order" can emerge at $T=T_{\rm cdw}$, depending on the fourth-order terms of Ginzburg-Landau free energy. 
The single-$\bm{q}$ (double-$\bm{q}$) bond-order possesses the $C_2$ ($C_4$) symmetry. 
In the main text, we focus on the electronic states given by the single-$\bm{q}$ bond-order. 
Below, we discuss the electronic states given by the double-$\bm{q}$ bond-order. 

\begin{figure}[!htb]
\includegraphics[width=.99\linewidth]{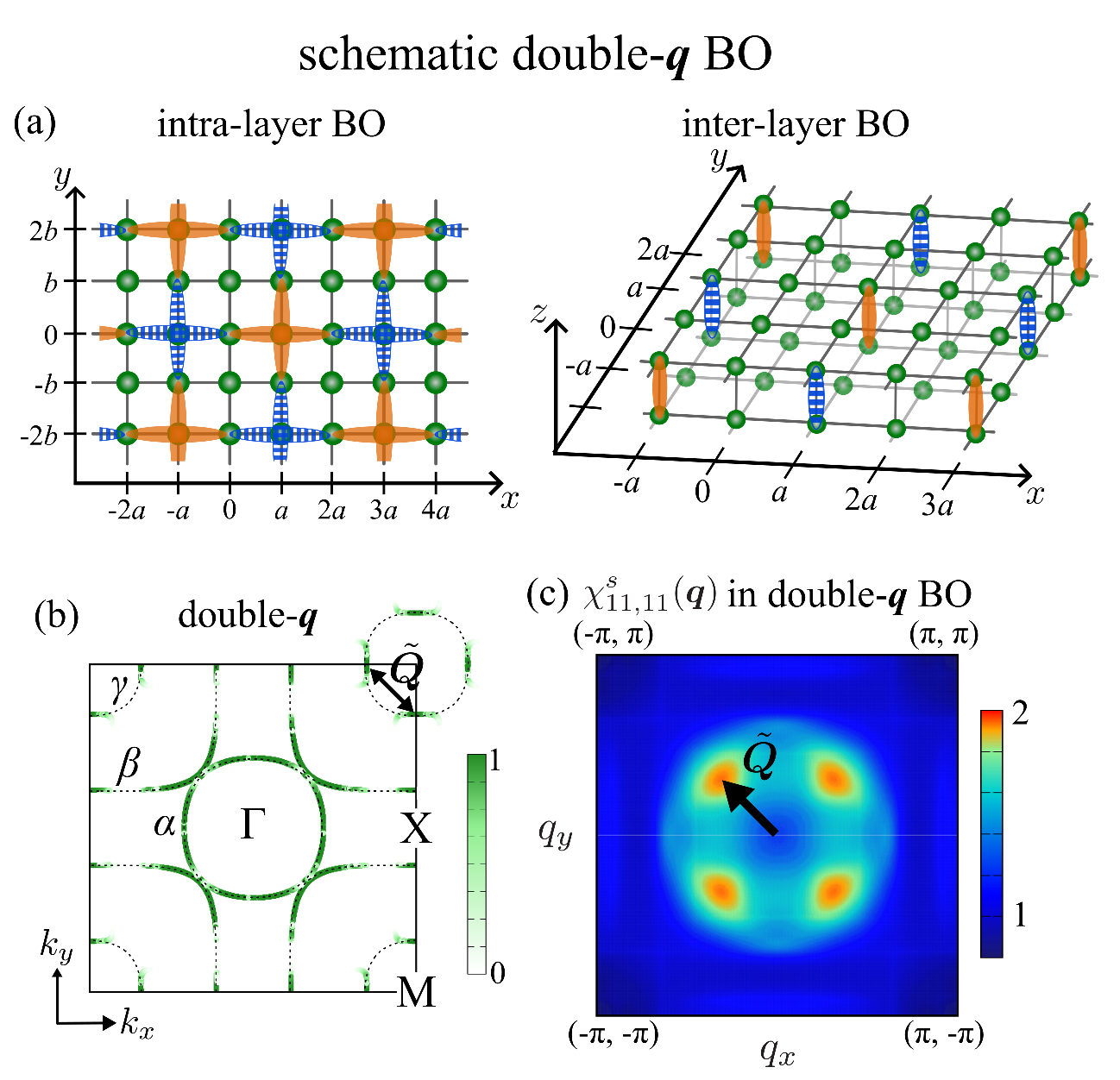}
\caption{
(a) Simple schematic picture of the double-$\bm{q}$ inter/intra-layer bond-order 
(coexistence of $\q=\bm{Q}_1$ CDW and $\q=\bm{Q}_2$ CDW) in real space. 
Both intra-layer bond-order and inter-layer bond-order are shown. 
Orange-filled (blue-striped) bonds represent the hopping integrals modulated by $+\delta t>0$ ($-\delta t <0$). 
Here, we denote only the vertical and the third-nearest bond modulations to simplify the picture. 
(b) Unfolded FSs under the double-$\bm{q}$ bond-order (green line), and original FSs (black dotted lines). 
(c) Spin susceptibility $\chi^s_{11,11}(\bm{q})$ under the double-$\bm{q}$ bond-order, by setting $U=1.0$ eV and $f^{\rm max}=0.02$ eV. 
Here, $\tilde{\bm Q}\approx(\pm0.4 \pi, \pm 0.4\pi)$ is the nesting vector due to the remaining $\gamma$-pocket with electron-doping. 
Note that $|\tilde{\bm Q}|\lesssim|{\bm Q}|$.
}
\label{fig:figA3}
\end{figure}

Figure \ref{fig:figA3} (a) exhibits a simplified picture of double-$\bm{q}$ inter/intra-layer bond-order state in real space, derived from the form factor $\tilde{F}^{ll'}(\bm{r})$ shown in Figs. \ref{fig:fig4} (e) and (f).
The schematic picture is derived from $\delta t_{ij}$, as given in Eq. (\ref{eqn:hopping_modulation}), through the summation over $\bm{q} = \bm{Q}_1$ and $\bm{Q}_2$. 
Here, we express only the third-nearest neighbor bond-order and set the phase as $\psi=\pi/2$. 
Likewise, in the case of the single-$\bm{q}$ bond-order, 
the local inversion symmetry is broken, which is understood by 
choosing the inversion center as $(x,y)=(0,0)$. 
On the other hand, the present CDW does not violate 
the global inversion symmetry, whose center is $(x,y)=(+a,0)$ ($C_4$ symmetry).

Figure \ref{fig:figA3} (b) shows the FSs under the double-$\bm{q}$ bond-order (green line), 
where the black dotted line is the original FSs. 
We put $f^{\mathrm{max}}=0.02$ eV. 
In the case of the double-$\bm{q}$ bond-order, the $\gamma$-pocket 
almost vanishes, while a wide range of the $\alpha$- and $\beta$-pockets remain. 
According to the ARPES experiment under 160 K \cite{SYang-Ni-ARPES},
the $\gamma$-pocket is invisible, while the $\alpha$, $\beta$-pockets are clearly observed.
Therefore, the disappearance of the $\gamma$-pocket due to the double-$\bm{q}$ 
bond-order is consistent with the ARPES measurement. 

Figure \ref{fig:figA3} (c) exhibits RPA spin susceptibility $\chi^s_{11,11}(\bm{q})$ 
under the double-$\bm{q}$ bond-order, derived from the unfolded Green function method. 
Here, we set $U=1.0$ eV and $f^{\rm max}=0.02$ eV. 
The maximum peak appears at $\bm{q}\approx \tilde{\bm Q}$, where $\tilde{\bm Q}\approx(\pm0.4 \pi, \pm 0.4\pi)$ is the nesting vector due to the remaining $\gamma$-pocket. 
Thus, under the double-$\bm{q}$ bond-order, 
$\chi^s$ takes maximum value at $\bm{q}\approx (\pm \pi/2, \pm \pi/2)$ 
by magnifying the $\gamma$-pocket via slight self-hole-doping.

Therefore, in both cases that single-$\bm{q}$ and double-$\bm{q}$ bond-order 
are realized, our theory can naturally explain the emergence of the CDW order (at $\q\approx\bm{Q}$) without magnetization, and 
double stripe CDW + SDW coexisting state following the CDW transition, 
which is suggested in Ref. \cite{SRen-Ni-X-ray}.


\section*{E: Origin of the long-range bond-order}

\begin{figure}[!t]
\includegraphics[width=.99\linewidth]{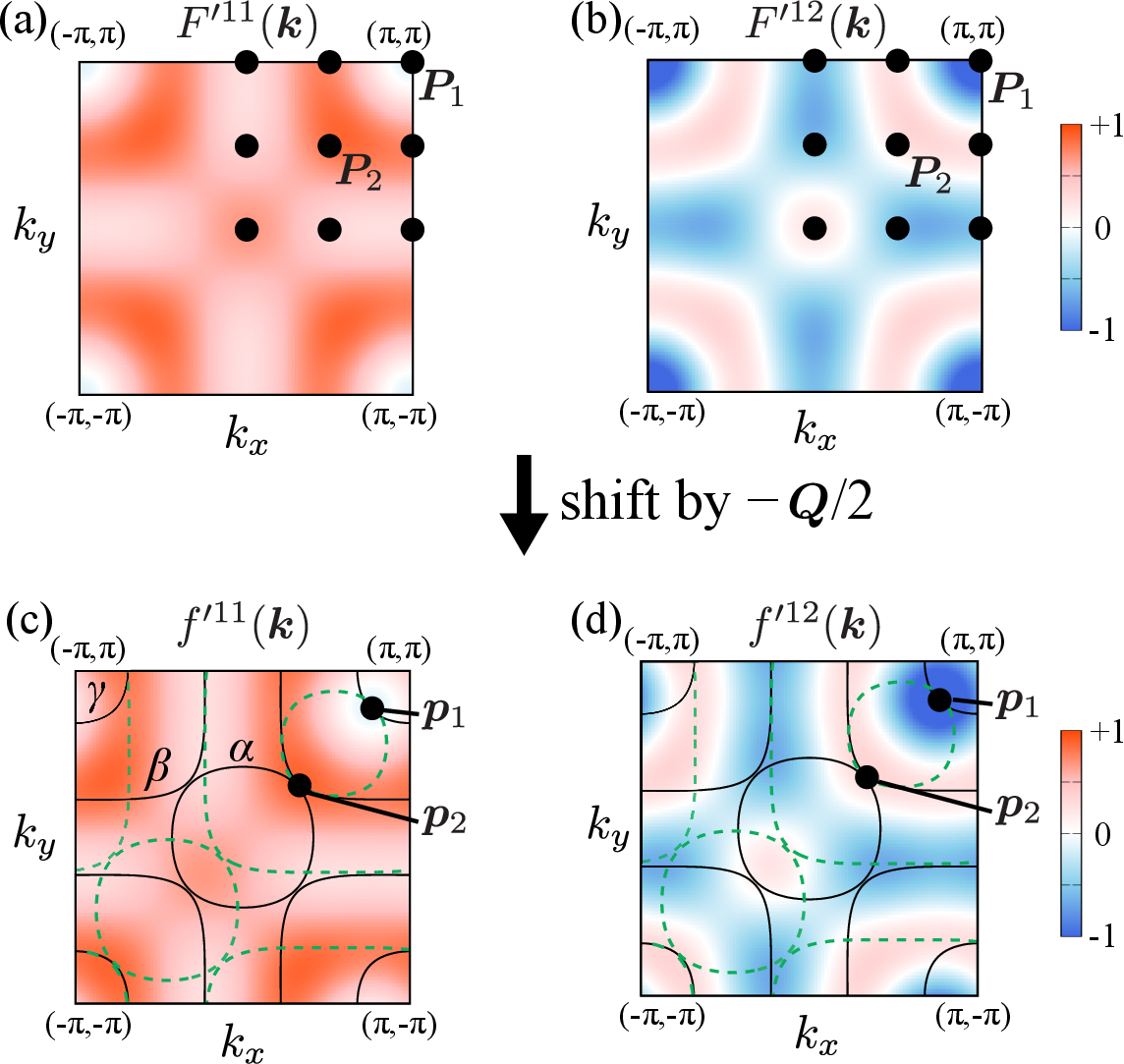}
\caption{
(a),(b) $\bm{k}$-dependences of $F^{\prime L}(\bm{k})$ ($L=11,12$) obtained by the Fourier transformation of $\tilde{F}^{\prime L}(\bm{r})$.
Here, $\tilde{F}^{\prime L}(\bm{r})$ given by Eq. (\ref{eqn:FTcheck}). 
For both $L=11$ and $12$, $F^{\prime L}(\bm{k})$ takes the minimum (maximum) value at $\bm{P}_1$ ($\bm{P}_2$).
Here, $\bm{P}_1 \equiv (\pi,\pi)$, $\bm{P}_2 \equiv (\pi/2,\pi/2)$. 
(c),(d) $\bm{k}$-dependences of $f^{\prime L}(\bm{k})= F^{\prime L}(\bm{k}+\bm{Q}/2)$. 
They are very similar to the original form factor $f_{\bm{Q}}^{L}$ shown in Fig. \ref{fig:fig4} (a) and (b) in the main text.
$f^{\prime L}(\bm{k})$ for $L=11$ ($L=12$) takes the minimum (maximum) value at $\bm{p}_1$ ($\bm{p}_2$).
Note that $\bm{P}_{1(2)}\equiv\bm{p}_{1(2)}+\bm{Q}/2$.
}
\label{fig:figA5}
\end{figure}

Here, we explain why the shifted real-space form factor $\tilde{F}^{L}(\bm{r})$ 
($L=11,12$) takes large values at the third- and fourth-nearest neighbors. 
We introduce a simplified shifted form factor 
$\tilde{F}^{\prime L} (\bm{r}) \equiv C(\bm{r}) \tilde{F}^{L}(\bm{r})$, 
where $C(\bm{r})=1$ for the origin ($\bm{r}={\bm0}$), third- and fourth-nearest neighbors 
($\bm{r}=(\pm2a,0), (0,\pm2b), (\pm2a\pm(\mp)b), (\pm a \pm (\mp) 2b)$), 
and $C(\bm{r})=0$ for others. 
Then, the Fourier transformation of $\tilde{F}^{\prime L}(\bm{r})$ is given as 
\begin{eqnarray}
	F^{\prime L}(\bm{k}) = A &+& B\sum_{s=\pm} \{\cos(s2k_x)+\cos(s2k_y)\} \notag \\
	&+& C\sum_{s,s'=\pm}\{\cos(s2k_x+s'k_y) \notag \notag \\
	&&\ \ \ \ \ \ \ \ \ \ \ \ \ \ \ \ +\cos(sk_x+s'2k_y)\}, \notag \\
	\label{eqn:FTcheck}
\end{eqnarray}
where $(A,B,C)=(+0.54, -0.07, +0.05)$ for $L=11$, and $(A,B,C)=(-0.18, -0.10, +0.10)$ for $L=12$. 
(Exactly speaking, $\tilde{F}^{L}(\bm{r}=(2a,b))\neq \tilde{F}^{L}(\bm{r}=(2a,-b))$ as shown in Figs. \ref{fig:fig4} (c) and (d) due to the $\bm{q}\neq \bm{0}$. 
However, we neglect this tiny difference to simplify the discussion of this section.) 
Figures. \ref{fig:figA5} (a) and (b) show $F^{\prime L}(\bm{k})$ given in Eq. (\ref{eqn:FTcheck}) for $L=11$ and $12$, respectively. 
For both $L=11$ and $12$, $F^{\prime L}(\bm{k})$ takes the minimum (maximum) value at $\bm{P}_1$ ($\bm{P}_2$).
Here, $\bm{P}_1 \equiv (\pi,\pi)$ and $\bm{P}_2 \equiv (\pi/2,\pi/2)$.

At black dots in Figs. \ref{fig:figA5} (a) and (b), $F^{\prime L}(\bm{k})$ takes the following values:
\begin{equation}
	\begin{cases}
	   A+4B+4C \ \ \ \ \ \ \ \ \ \ \mathrm{at}\ \bm{k}=(0,0)\\
	   A-4C \ \ \ \ \ \ \ \ \ \ \ \ \ \ \ \ \ \mathrm{at}\ \bm{k}=(0,\pi/2)=(\pi/2,0)\\
	   A+4B \ \ \ \ \ \ \ \ \ \ \ \ \ \ \ \ \ \mathrm{at}\ \bm{k}=(0,\pi)=(\pi,0)\\
	   A-4B \ \ \ \ \ \ \ \ \ \ \ \ \ \ \ \ \ \mathrm{at}\ \bm{k}=(\pi/2,\pi/2)\equiv \bm{P}_2\\
	   A+4C \ \ \ \ \ \ \ \ \ \ \ \ \ \ \ \ \ \mathrm{at}\ \bm{k}=(\pi,\pi/2)=(\pi/2,\pi)\\
	   A+4B-8C \ \ \ \ \ \ \ \ \ \ \mathrm{at}\ \bm{k}=(\pi,\pi) \equiv \bm{P}_1
	\end{cases}
 \end{equation}
Here, we draw $f^{\prime L}(\bm{k})\equiv F^{\prime L}(\bm{k}+\bm{Q}/2)$
in Figs. \ref{fig:figA5} (c) and (d). 
For both $L=11$ and $12$, $f^{\prime L}(\bm{k})$ takes the minimum (maximum) value at $\bm{p}_1$ ($\bm{p}_2$).
Note that $\bm{P}_{1(2)}\equiv\bm{p}_{1(2)}+\bm{Q}/2$.
They are very similar to the original form factor $f_{\bm{Q}}^{L}$ shown in 
\ref{fig:fig4} (a) and (b) in the main text.
Therefore, the form factor mainly originates from the hopping modulations between the third- and fourth-nearest neighbor sites.

\section*{F: Pairing interaction $V^{\mathrm{SC}}(\theta,\theta')$ derived from the Bethe-Salpeter equation method}

Here, we explain the $\bm{k}$, $\bm{k}'$-dependences of pairing interaction mediated by CDW and SDW fluctuations. 
\begin{figure}[!htb]
	\includegraphics[width=.75\linewidth]{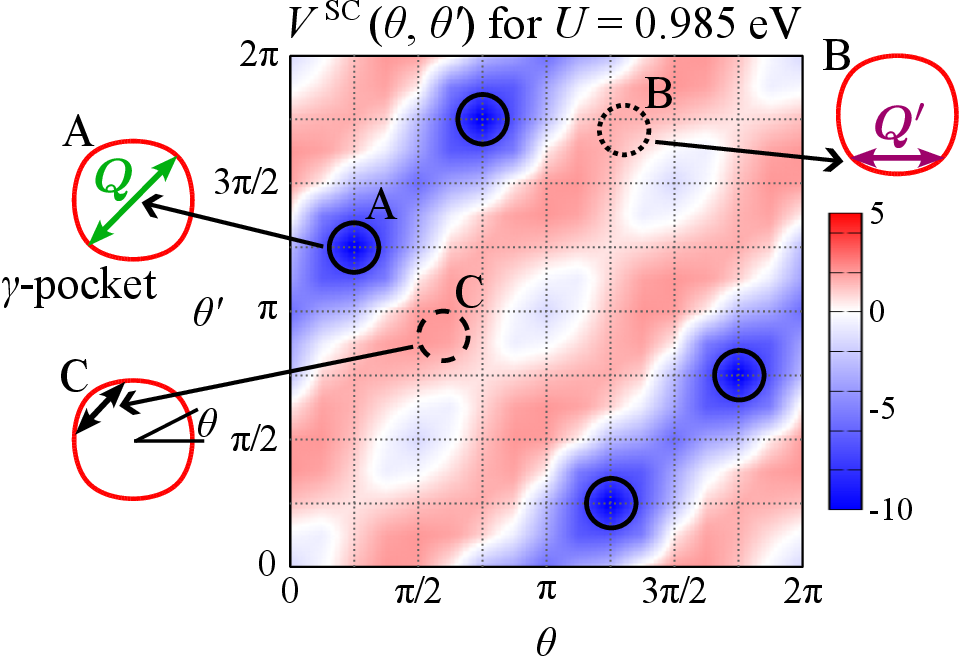}
	\caption{
	Pairing interaction $V^{\mathrm{SC}}(\theta,\theta')$ due to bond-order and spin fluctuations. 
	$\theta=\arctan(k_y/k_x)$ and $\theta'=\arctan(k_y'/k_x')$. 
	The blue (red) area represents attractive (repulsive) pairing interaction induced by bond-order fluctuations (spin fluctuations). 
	The circle A and B indicate $V^{\mathrm{SC}}$ between Fermi momenta corresponded to the nesting vectors $\bm{Q}$ and $\bm{Q}'$, respectively. 
	Also, the circle C indicates the largest repulsive pairing interaction.
	}
	\label{fig:V_SC_BS}
	\end{figure}
In the main text, we discussed the bond-order and spin fluctuations-mediated pairing interaction leads to 
$d_{xy}$-wave and $s_{\pm}$-wave SC states. 
The pairing interaction $V^{\mathrm{SC}}(\theta,\theta')$ for $U=0.985$ eV is exhibited in Fig. \ref{fig:V_SC_BS}. 
Here, $\theta=\arctan(k_y/k_x)$ and $\theta'=\arctan(k_y'/k_x')$. 
We depict the pairing interaction between Fermi momenta on the $\gamma$-pocket because it is much larger compared to other interactions. 
The red area shows the repulsive interaction induced by spin fluctuations, and the blue area shows the attractive interaction induced by bond-order fluctuations $V_{\mathrm{bond}}$. 
The obtained bond-order fluctuation-mediated pairing interaction is more than twice as large as the spin fluctuation-mediated pairing interaction. 
The largest negative value of $V_{\mathrm{bond}}$ is $\sim -10$, which is consistent with $-f_{\bm{k}-\bm{k}'}(\bm{k}')f_{\bm{k}'-\bm{k}}(-\bm{k}') \frac{g_{\bm{k}-\bm{k}'}}{1-\lambda_{\bm{Q}}}$, 
where $\lambda_{\bm{Q}}=0.93$, and $f_{\bm{k}-\bm{k}'}(\bm{k}')=f_{\bm{k}'-\bm{k}}(-\bm{k}')=1$. 
Here, a value of $g_{\bm{k}-\bm{k}'}\sim U\sim 1$, which is realized in kagome metals reported in Ref. \cite{Tazai-kagome1}. 
In Fig. \ref{fig:V_SC_BS}, the circle A represents the strong attractive pairing interaction $V^{\mathrm{SC}}(k,k')$, where $\bm{k}-\bm{k}'=\bm{Q}$, which corresponds to the peak position of $\lambda_{\bm{q}}$ as shown in Fig. \ref{fig:fig3} (c).
This interaction favors both $s$-wave and $d_{xy}$-wave SC states. 

Also, the circle B and C show the repulsive interaction induced by spin fluctuations. 
The circle B represents $V^{\mathrm{SC}}(k,k')$, where $\bm{k}-\bm{k}'=\bm{Q}'$, which corresponds to the peak position of $\chi^s({\bm{q}})$ as shown in Fig. \ref{fig:fig3} (a). 
This interaction favors $d_{xy}$-wave SC states as shown in Fig. \ref{fig:fig6} (c). 
Also, the repulsive interaction induced by $\chi^s({\bm{Q}'})$ between $\beta$- and $\gamma$-pockets (about half value of $V^{\mathrm{SC}}$ at the circle B) 
favors $s_{\pm}$-wave SC state as shown in Fig. \ref{fig:fig6} (d). 
The strongest repulsive interaction, indicated by the circle C, favors the $d_{x^2-y^2}$-wave SC state. 
This state is further stabilized by $V_{\mathrm{bond}}$ as shown in Fig. \ref{fig:fig6} (a), while its stability is constrained by the nodal structure of the $d_{x^2-y^2}$-wave gap. 

We note that both the bond-order and spin fluctuations-mediated pairing interactions 
enhance $T_c$, and favor $s_{\pm}$-wave and $d_{xy}$-wave SC states cooperatively in the present theory.


\end{document}